\documentclass[12pt,a4paper]{article}
\pdfoutput=1
\usepackage{macros}

\def\cY{{\mathcal{Y}}}

\preprint{}
\title{ABCD of qq-characters}

\author[1]{Satoshi Nawata,}
\author[2]{Kilar Zhang,}
\author[3]{Rui-Dong Zhu}
\affiliation[1]{Department of Physics and Center for Field Theory and Particle Physics, Fudan University, 2005, Songhu Road, 200438 Shanghai, China}
\affiliation[2]{Department of Physics, Shanghai University, Shanghai 200444, China\\ Shanghai Key Lab for Astrophysics, Shanghai 200234, China}
\affiliation[3]{Institute for Advanced Study \& School of Physical Science and Technology,\\ Soochow University, Suzhou 215006, China
}
\emailAdd{snawata@gmail.com}
\emailAdd{kilar@shu.edu.cn}
\emailAdd{rdzhu@suda.edu.cn}
\abstract{The qq-characters are powerful tools to reveal symmetries and integrabilities of Seiberg-Witten theories. The goal of this paper is to provide analytic expressions of qq-characters based on Young diagrams in 5d $\cN=1$ pure Yang-Mills theories with $BCD$-type gauge groups, by focusing on the unrefined limit. Using these expressions, we investigate the relationships among qq-characters of classical gauge groups. For $\SO(n)$ gauge groups, we construct a quantum-toroidal-like algebra via the Ward-identity approach, which allows us to derive the qq-characters. }

\begin{document}
\Yboxdim5pt

\allowdisplaybreaks

\maketitle

\section{Introduction}

The integrability structure in supersymmetric gauge theories has been intensively studied since the Seiberg-Witten exact result \cite{Seiberg:1994rs,Seiberg:1994aj} and its relation to classical integrable models \cite{Gorsky:1995zq,Martinec:1995by,Donagi:1995cf}. Considering supersymmetric theories on the $\Omega$-background, Nekrasov developed the technique of supersymmetric localization, and reformulated the Seiberg-Witten theory in terms of instanton partition functions \cite{NekrasovInstanton}. These developments have revealed that the BPS sectors of supersymmetric gauge theories on general $\Omega$-backgrounds are governed by algebras equipped with universal R-matrices, such as the quantum toroidal algebras and the affine Yangian algebras \cite{Maulik:2012wi,SHc,Awata:2011ce}. This hints at a deep connection between the integrability and supersymmetric gauge theories, potentially arising from some string-theoretical construction of the gauge theories. The algebraic formulation of supersymmetric gauge theories has been generalized to various kinds of gauge theories ranging from 2d $\cN=(2,2)$ theories to 6d $\cN=(1,0)$ theories (e.g. \cite{Zhu:2017ysu,Zenkevich:2018fzl,Bourgine:2018uod,Ghoneim:2020sqi,Zenkevich:2020ufs,Bourgine:2021nyw,Bourgine:2022scz}). However, the criteria for determining whether such integrable structures exist in a given theory is still not well understood, and even for pure gauge theories with $BCD$-type gauge groups, we lack a clear understanding of their integrable structure. One of the obstacles in studying $BCD$-type gauge theories is the complicated pole structure obtained from the Jeffery-Kirwan (JK) prescription \cite{Hwang:2014uwa,Nakamura:2014nha,Nakamura:2015zsa}. However, in the unrefined limit, the calculation is greatly simplified, as one can effectively label the JK poles with tuples of Young diagrams \cite{Hayashi:2020hhb,Nawata:2021dlk,Chen:2023smd}. This may provide a new avenue for investigating the integrability structures of $BCD$-type gauge theories.

Among various different formulations of the integrability in supersymmetric gauge theories, the qq-characters, a family of physical quantities first named in \cite{Nekrasov:2015wsu}, serve as a probe to the hierarchy of the integrable structure of gauge theories in different parameter regions. In the case of $A$-type gauge groups, the qq-characters as operator-valued quantities generate a so-called quiver $\cW$-algebra, that is the $\cW$-algebra associated to the Dynkin diagram corresponding to the quiver structure of the gauge theory \cite{Kimura:2015rgi,Kimura:2016dys}. In the Nekrasov-Shatashvili limit, which turns off one $\Omega$-background parameter, its expectation value gives the TQ-relation of a class of quantum integrable models \cite{Nekrasov:2013xda}, and further in the classical limit, the expectation value reduces to the Seiberg-Witten curve of supersymmetric gauge theories with eight supercharges. In supersymmetric gauge theories with $BCD$-type gauge groups, the brane construction of qq-characters was proposed in \cite{Haouzi:2020yxy,Haouzi:2020zls} as a generalization of \cite{Kim:2016qqs}. Finding exact forms of qq-characters in 5d $\cN=1$ supersymmetric gauge theories with $BCD$-type gauge groups has been a challenging task due to the complexity of the infinite number of terms of the natural quantity called the $Y$-function and the evaluation of JK residues. In this paper, we address this issue by focusing on the unrefined limit, which greatly simplifies the computation, and provide analytic expression of qq-characters for these theories. This will provide further strong evidence to support the proposal of analytic expression as a summation over Young diagrams for the instanton partition functions presented in \cite{Nawata:2021dlk}. Furthermore, using the analytic expression of qq-characters, we explore the algebraic structure in gauge theories that may connect to the integrability.

This paper is organized as follows. In \S\ref{s:qq-unref}, we review the concept of qq-characters as co-dimension four defect partition functions and provide an analytic expression based on Young-diagram summation for classical gauge groups. In \S\ref{s:group}, we examine the Lie-algebraic relations between qq-characters for isomorphic gauge algebras, which serves as a validation of the analytic expressions derived in \S\ref{s:qq-unref}.  In \S\ref{s:qq-alg}, we delve into the construction of a quantum-toroidal-like algebra that reproduces the expression of the (fundamental) qq-character for $\SO(n)$ gauge theories, utilizing the Ward identity approach outlined in \cite{BMZ,5dBMZ}. Additionally, several appendices are included to supplement the main text and to provide specific computational details.

\section{ABCD of qq-characters}\label{s:qq-unref}
In this section, we will give an overview of the method for deriving qq-characters through localization computation and present ``finite'' expressions for the unrefined limit of the qq-characters for classical gauge groups, evaluated in terms of Young diagrams. This will provide a clear interpretation of the qq-characters as the quantization of Seiberg-Witten curves.

\subsection{qq-characters as defect partition functions}\label{sec:qq}

The concept of the qq-characters was first introduced by Nekrasov in the context of BPS/CFT correspondence \cite{Nekrasov:2015wsu}, which generalizes the q-characters of quantum affine algebras \cite{Frenkel:1998ojj}. They can also be understood as partition functions of 4d/5d supersymmetric gauge theories with co-dimension four defects on the $\Omega$-background \cite{Kim:2016qqs,Haouzi:2020yxy}. In this paper, we will focus specifically on the case of pure Yang-Mills theory with 8 supercharges. The $\Omega$-background effectively reduces the 5d $\cN=1$ pure Yang-Mills theory into supersymmetric quantum mechanics on the instanton moduli spaces. Since the ADHM descriptions of the instanton moduli spaces are known for classical gauge groups $G$, the partition function of pure Yang-Mills theory in the presence of co-dimension four defects can be expressed by a Jeffery-Kirwan (JK) residue integral:
\begin{equation}\label{defect}
Z_{\textrm{defect}}^\frakg(\vec{z})=\sum_{k=0}^\infty \frac{\mathfrak{q}^k}{|W(G_k)|}\oint_{\textrm{JK}}\prod_{i=1}^k\frac{\textrm{d}\phi_i}{2\pi i} Z^{(k)}_{\textrm{vec}}Z^{(k)}_{\textrm{def}}(\vec{z})~,
\end{equation}
where $\mathfrak{q}$ indicates the gauge coupling constant (or the dynamical scale after taking the decoupling limit). Note that $G_k$ is the gauge group  of supersymmetric quantum mechanics on the $k$-instanton moduli space, and $|W(G_k)|$ is the order of its Weyl group.
 The JK residue integral \cite{jeffrey1995localization,brion1999arrangement,Benini:2013xpa} is performed on the gauge fugacities  $\phi_i$ of the instanton quantum mechanics. The contribution $Z^{(k)}_{\textrm{vec}}$ to the integrand can be obtained from the ADHM description as given in \cite{NekrasovInstanton,ABCD-instanton,Kim:2012gu}. To read off the defect contribution $Z^{(k)}_{\textrm{def}}(\vec{z})$, one can consider the brane configurations and open string spectra between D-branes \cite{Kim:2016qqs,Haouzi:2020yxy}.
 
\begin{table}
\begin{center}
\begin{tabular}{|c|c|c|c|c|c|c|c|c|c|c|}
\hline
& 0 & 1 & 2 & 3 & 4& 5 & 6 & 7 & 8 & 9 \cr
\hline
D4/O4 & $\bullet$ & $\bullet$ & $\bullet$ & $\bullet$ & $\bullet$ & $-$ & $-$ & $-$ & $-$ & $-$ \cr
\hline
D0 & $\bullet$ & $-$ & $-$ & $-$ & $-$ & $-$ & $-$ & $-$ & $-$ & $-$ \cr
\hline
D$4'$ & $\bullet$ & $-$ & $-$ & $-$ & $-$ & $\bullet$  & $\bullet$ & $\bullet$ & $\bullet$ & $-$ \cr
\hline
\end{tabular}
\caption{Configuration of branes in the brane web construction of ADHM construction with Wilson lines (introduced by D4' branes). }
\label{fig:brane-setup}
\end{center}
\end{table}

Let us consider the brane configuration given by Figure \ref{fig:brane-setup} in Type IIA theory. In addition, we introduce the $\Omega$-deformation where $\U(1)_{\e_1}\times \U(1)_{\e_2}$ rotates the $x^{1234}$-plane $\bR^4$ while $\U(1)_{m} \times \U(1)_{-m}$ rotates the $x^{5678}$-plane $\bR^4$. Consequently, the worldvolume theory on a stack of D4-branes is 5d $\cN=1^*$ theory where the adjoint hypermultiplet has mass $m$. The correspondence between the type of 5d gauge group and the kind of O4-plane is as follows:
$$\begin{array}{ccc}
A &\qquad & \mathrm{No}\\
B &\qquad & \widetilde{\mathrm{O} 4}^{-} \\
C &\qquad &\mathrm{O} 4^{+},  \widetilde{\mathrm{O} 4}^{+} \\
D &\qquad & {\mathrm{O} 4}^{-} \\
\end{array}$$
D0-branes give rise to instantons and the D4'-brane realizes the co-dimension four defect. As we will see, the defect can be understood as Wilson loops. Recall that the D0-branes give rise to the gauge group $G_k$ and the D4-branes to the flavor symmetry $G$ in the $\cN=4$ supersymmetric quantum mechanics on the instanton moduli space.  From this viewpoint, additional fields emerge from the string connections between D0- and D4'-branes, as well as the strings between D4- and D4'-branes. Table~\ref{tab:field-content} provides a list of these fields. Specifically, the hypermultiplet and Fermi multiplet charged under $G_k$ arise from the D0-D4' strings while the Fermi multiplet charged under $G$ stems from the D4-D4' strings. These multiplets have a real mass parameter $\zeta$ associated to the relative distance of D4- and D4'-branes along $x^9$-direction. 
\begin{table}[!h]
\centering
\begin{tabular}{|c|c|c|c|}
\hline
Multiplet  & $G_k$ & $G$ \\
\hline 
Hyper  & $\square$ & $\emptyset$ \\
\hline
Fermi & $\square$ &  $\emptyset$ \\
\hline
Fermi  &  $\emptyset$ & $\square$ \\
\hline
\end{tabular}
\caption{Additional multiplets due to the presence of the D4'-brane.}
\label{tab:field-content}
\end{table}

By taking the decoupling limit $m\to\infty$, we can obtain the partition function of the pure Yang-Mills theory with the presence of the defect. This brane configuration is considered in \cite{Kim:2016qqs} for $G=\SU(N)$ while O-planes of other types are introduced in \cite{Haouzi:2020yxy} for the other classical gauge groups. Nevertheless, as we will see below, the defect contributions become the same as in \cite{Haouzi:2020yxy} for the pure Yang-Mills. In what follows, we provide the explicit expressions for the defect partition functions of 5d theories with classical gauge groups.

Throughout the article, we use $q_1=e^{-\e_1}$, $q_2=e^{-\e_2}$ for the parameters of the $\Omega$-deformation, and we use the notation $\epsilon_\pm=\frac{\epsilon_1\pm\epsilon_2}{2}$. The unrefined limit refers to $\e_1=-\e_2=\hbar$ so that the 5d parameter is $q=e^{-\hbar}=q_1=q_2^{-1}$.

\subsubsection*{Type $A$}
To begin with, we will review the well-established case of gauge groups of $A$-type. In this case, $G_k=\U(k)$ and the integrand of the instanton partition function is
\begin{equation}
Z^{(k)}_{\textrm{vec}}=e^{\kappa \phi_i}\lt(\frac{\llbracket2\epsilon_+\rrbracket}{\llbracket \epsilon_1\rrbracket\llbracket \epsilon_2\rrbracket}\rt)^k\prod_{i=1}^k\prod_{\alpha=1}^N\frac{1}{\llbracket \epsilon_+\pm (\phi_i-\mathfrak{a}_\alpha)\rrbracket}\prod_{\substack{i,j=1\\i\neq j}}^k\cS(\phi_i-\phi_j+\epsilon_+)^{-1}~.\label{int-Zvec}
\end{equation}
From Figure \ref{tab:field-content}, one can read off the defect contribution \cite{Kim:2016qqs}
\begin{equation}
Z^{(k)}_{\textrm{def}}(\vec{z})=\prod_{j=1}^{w}\prod_{\alpha=1}^N\llbracket \zeta_j-\mathfrak{a}_\alpha+2\e_+\rrbracket\prod_{i=1}^k\cS(\zeta_j-\phi_i),\label{A-def}
\end{equation}
with $z_j=\exp\lt(-\zeta_j\rt)$, $A_\alpha=\exp\lt(-\mathfrak{a}_\alpha\rt)$ and $\cS$ is given by
\begin{equation}\label{S}
\cS(\zeta):=\frac{\llbracket \epsilon_-\pm \zeta\rrbracket}{\llbracket\epsilon_+\pm  \zeta\rrbracket}~.
\end{equation}
In this paper, we adopt the convention that $\llbracket \alpha\rrbracket:=2\sinh\left(\frac{\alpha}{2}\right)$ and  $\llbracket \alpha \pm \beta \rrbracket := \llbracket \alpha + \beta \rrbracket \llbracket \alpha - \beta \rrbracket$.
Note that $\kappa$ in \eqref{int-Zvec} is the 5d Chern-Simons level, and it is set to be zero except for \S\ref{sec:sp1-su2}. Here, the positions of co-dimension four defects are denoted by $\zeta_j$, with the total number $w$ (namely the number of D4'-branes). However, our primary focus is on the case of $w=1$, unless otherwise specified (only in \eqref{qq-A-w2}).

In the case of $A$-type, the qq-character was obtained with generic $\Omega$-background parameter $(\e_1,\e_2)$ \cite{BPS/CFT}. This is possible because fixed point sets of the equivariant actions on the $k$-instanton moduli spaces are classified by $N$-tuples $\vec{\lambda}$ of Young diagrams with the total number of boxes $|\Vec{\lambda}|=k$, and the instanton partition function can be expressed as a sum over $N$-tuples $\vec{\lambda}$ of Young diagrams at the refined level \cite{NekrasovInstanton}. Also, by performing the JK residue integrals at the refined level, one can obtain the same result \cite{Hwang:2014uwa}. When there are no defects (i.e., when $w=0$), the instanton partition function is therefore given by:
\begin{equation}\label{instnaton-A}
Z_{\textrm{inst}}^{\mathfrak{su}(N)}=\sum_{\vec\lambda}\mathfrak{q}^{|\vec{\lambda}|}Z^{\fraksu(N)}_{\vec{\lambda}}~, \qquad  Z^{\fraksu(N)}_{\vec{\lambda}}=\prod_{\alpha,\beta=1}^N N^{-1}_{\lambda^{(\alpha)}\lambda^{(\beta)}}(Q_{\alpha\beta},q_1,q_2),
\end{equation}
where the Coulomb branch parameters enter the formula via $Q_{\alpha\beta}=A_\alpha/A_\beta$, and  the Nekrasov factor $N_{\lambda\nu}(Q,q_1,q_2)$ is expressed by
\begin{equation}
N_{\lambda\nu}(Q,q_1,q_2):=\prod_{(i,j)\in \lambda}\lt(1-Qq_2^{-\nu_i+j}q_1^{\lambda^t_j-i+1}\rt)\prod_{(i,j)\in \nu}\lt(1-Qq_2^{\lambda_i-j+1}q_1^{-\nu^t_j+i}\rt).\label{def:Nekra}
\end{equation}
Some relevant properties of the Nekrasov factor are listed in Appendix \ref{a:id}. Here, for later use, we write a fixed point set of the equivariant actions $\U(1)_{\e_{1,2}}\times \U(1)_{\vec{\fraka}}$ (or equivalently the sets of JK poles)   as ${\ket{\vec{A},\vec{\lambda}}}_{\textrm{ref}}$, and represent the instanton partition function with the inner product of Gaiotto states \cite{Gaiotto:2009ma}
\begin{equation}\label{Gaiotto}
\ket{\frakG}=\sum_{\vec{\lambda}}\lt(\mathfrak{q}^{|\vec\lambda|}Z^{\fraksu(N)}_{\vec{\lambda}}\rt)^{\frac12}\ket{\vec{A},\vec{\lambda}}_{\textrm{ref}},\qquad \langle \frakG\ket{\frakG}=Z_{\textrm{inst}}^{\mathfrak{su}(N)}~.
\end{equation}
Recall that the pure Yang-Mills theory arises from a sphere with two irregular punctures in the class $\cS$ construction \cite{Gaiotto:2009ma}, and the irregular puncture corresponds to the Gaiotto state as a coherent state in the dual $\cW$-algebra \cite{Matsuo:2014rba}.

As done in  \cite{Kim:2016qqs}, even with one defect $w=1$, the JK poles are also classified by two types of $N$-tuples of Young diagrams. Poles in the one type come from $Z_{\mathrm{vec}}^{(k)}$, and therefore they are the same as $\vec{\lambda}$  without the defect. On the other hand, in the other type, one of the poles is located at $\phi_i=\zeta-\e_+$, which comes from $Z_{\textrm{def}}^{(k)}$, and the other poles originate from $Z_{\mathrm{vec}}^{(k)}$, which are classified by $\vec{\lambda'}$ where $|\vec{\lambda}^{\prime}|=|\vec{\lambda}|-1$. The contributions from the poles of these two types can be nicely packaged into two expectation values as
\begin{equation}
\langle\chi^A(z)\rangle=Z_{\textrm{defect}}^{\mathfrak{su}(N)}(z)=\langle Y^A(z)\rangle +\left\langle \frac{\mathfrak{q}c_{A}}{Y^A(zq_3)}\right\rangle,\label{qq-A-1}
\end{equation}
where $c_{A}=(-1)^N$ and $q_3= (q_1q_2)^{-1}$. Here, $\langle Y^A(z)\rangle$ can be interpreted as the expectation value of the defect \eqref{A-def} where
the expectation value of an operator $\cO$ in the 5d pure Yang-Mills theory is defined by sandwiching with the Gaiotto state $\ket{\frakG}$
\begin{equation}
    \langle \cO\rangle:=\bra{\frakG}\cO\ket{\frakG}~.
\end{equation}
and we can interpret that an operator $Y^A(z)$ acts diagonally on the basis $\{\ket{\vec{A},\vec{\lambda}}_{\textrm{ref}}\}$ as
\begin{equation}\label{YA}
    Y^A(z)\ket{\vec{A},\vec{\lambda}}_{\textrm{ref}}=\cY^A_{\vec{\lambda}}(z)\ket{\vec{A},\vec{\lambda}}_{\textrm{ref}},
\end{equation}
with
\begin{align}
\cY^A_{\vec{\lambda}}(z)=\prod_{\alpha=1}^N\llbracket \zeta-\mathfrak{a}_\alpha +2\e_+\rrbracket\prod_{x\in\vec{\lambda}}\cS(\zeta-\phi_x)
=\frac{\prod_{x\in \frakA(\vec{\lambda})}\llbracket \zeta-\phi_x+\epsilon_+\rrbracket}{\prod_{x\in \frakR(\vec{\lambda})}\llbracket \zeta-\phi_x-\epsilon_+\rrbracket},\label{def-YA}
\end{align}
and
\be \label{pole-location}
\phi_x =\mathfrak{a}_\alpha-\e_++(i-1) \epsilon_1+(j-1) \epsilon_2~, \qquad x=(i,j) \in \lambda^{(\alpha)}
\ee
Note that $\frakA(\Vec{\lambda})$/$\frakR(\Vec{\lambda})$ is the set of boxes in $\Vec{\lambda}$ that can be added/removed, respectively.
Consequently, the expectation value of $Y^A(z)$ is then given by
\begin{equation}
\langle Y^A(z)\rangle:=\sum_{\vec{\lambda}}\mathfrak{q}^{|\vec{\lambda}|}\cY^A_{\vec{\lambda}}(z)Z^{\fraksu(N)}_{\vec{\lambda}}.\label{exp-YA}
\end{equation}
There exist unphysical poles in both $\cY^A_{\vec{\lambda}}(z)$ and $1/\cY^A_{\vec{\lambda}}(zq_3)$. However, it can be shown that their residues are zero through the evaluation of the JK residue integral because the pole in $\cY^A_{\vec{\lambda}}(z)$ cancels out with the pole in ${c_A}/{\cY^A_{\vec{\lambda}'}(zq_3)}$, where $\vec{\lambda}'$ satisfies $|\vec{\lambda}'|=|\vec{\lambda}|-1$. Additionally, the asymptotic behavior of the qq-character near $z\sim 0$ and $z\sim \infty$ can be expressed as:
\begin{align}
    &z^{\frac{N}{2}}\lt\langle\chi(z)\rt\rangle\sim z^N,\quad z\sim\infty\cr
    &z^{\frac{N}{2}}\lt\langle\chi(z)\rt\rangle\sim \cO(z^0),\quad z\sim 0,\label{asym-qq-A}
\end{align}
which implies that $z^{\frac{N}{2}}\lt\langle\chi(z)\rt\rangle$ is a degree-$N$ polynomial of $z$, i.e.
\begin{equation}
\lt\langle\chi(z)\rt\rangle=z^{-\frac{N}{2}}\sum_{i=0}^Nf_iz^i~.
\end{equation}
The coefficients $f_i$ are rational functions of $\vec{A}$, $q_{1,2}$. For further information on qq-characters of gauge theories with $A$-type gauge group, we refer to \cite{Kimura:2016ebq} for a comprehensive review.

As suggested in \eqref{qq-A-1}, the qq-characters actually have an algebraic meaning of the quiver structure of $A$-type gauge theories (at the classical level, this relation between the Seiberg-Witten curve and the fundamental character of quiver was first noticed in \cite{Nekrasov:2012xe} and was promoted to the full $\Omega$-deformed region in \cite{Nekrasov:2015wsu,Kimura:2015rgi}). For example, the character of the fundamental representation of $\fraksu(2)$ is
\begin{equation}
    \chi=y+y^{-1},
\end{equation}
and the corresponding qq-character can be understood as a doubly-quantized version of the Seiberg-Witten curve. The character for $w=2$ corresponds to that of the adjoint representation
\begin{equation}
    \chi_{w=2}=y^2+1+y^{-2},
\end{equation}
and is lifted to
\begin{align}
    \langle\chi_{w=2}(\vec{z})\rangle=&\langle Y^A(z_1)Y^A(z_2)\rangle+\mathfrak{q}c_A\lt\langle\cS(\zeta_1-\zeta_2+\epsilon_+)\frac{Y^A(z_1)}{Y^A(z_2q_3)}\rt\rangle \label{qq-A-w2}\\
    &+\mathfrak{q}c_A\lt\langle\cS(\zeta_2-\zeta_1+\epsilon_+)\frac{Y^A(z_2)}{Y^A(z_1q_3)}\rt\rangle+\mathfrak{q}^2c^2_A\lt\langle\frac{1}{Y^A(z_1q_3)Y^A(z_2q_3)}\rt\rangle.\nonumber
\end{align}
This interesting relation was found in \cite{Nekrasov:2012xe} for $A$-type gauge theories with $ADE$-quiver structures.

\subsubsection*{Type $BD$}

The nature of qq-characters has been well investigated for $A$-type gauge groups, but little has been done for other types. Let us consider the case $G=\SO(n)$ in which the instanton quantum mechanics is described by $G_k=\Sp(k)$.

Writing $n=2N+\xi$  ($n\equiv\xi \bmod 2$),  the integrand of the instanton partition function for the SO($n$) puer Yang-Mills \cite{ABCD-instanton} is given by
\begin{align}
    Z^{(k)}_{\textrm{vec}}=&\frac{\llbracket 2\epsilon_+\rrbracket^k}{\llbracket \epsilon_{1,2}\rrbracket^k}  \prod_{i=1}^k \frac{\llbracket \pm 2\phi_i\rrbracket\llbracket 2\epsilon_+\pm 2\phi_i\rrbracket}{\llbracket \epsilon_+ \pm\phi_i\rrbracket^{\xi}\prod_{\alpha=1}^N\llbracket \epsilon_+ \pm \phi_i\pm \mathfrak{a}_\alpha\rrbracket} \prod_{i<j}\cS(\epsilon_+\pm \phi_i\pm\phi_j)^{-1}
~.\label{int-ZOe}
\end{align}
The contribution from the defect can be derived from Table \ref{tab:field-content}
\begin{equation}\label{SO-def}
   Z^{(k)}_{\textrm{def}}(z)=\llbracket \zeta\rrbracket^\xi\prod_{\alpha=1}^N\llbracket \zeta\pm \mathfrak{a}_\alpha\rrbracket\prod_{i=1}^k\cS(\zeta\pm\phi_i)~,
\end{equation}
where $\cS$ is defined in \eqref{S}. This is equivalent to that derived in \cite{Haouzi:2020yxy}. 
The integration is carried out using the JK prescription as usual.

\subsubsection*{Type $C$}

In Sp($N$) theories, the calculation of the defect partition function is more involved because of the discrete $\theta$-angle \cite{Bergman:2013ala}. In fact, the gauge group of the Sp($N$) instanton quantum mechanics is $G_k=\OO(k)$ \cite{ABCD-instanton,Kim:2012gu}, which has two connected components $\OO(k)^\pm$. A choice of taking the sum or difference of these two contributions depends on the $\theta$-angle in the theory:
\begin{equation}\label{C-defect-PF}
Z_{\textrm{defect}}^{\mathfrak{sp}(N)_\theta}(z)=\sum_{k=0}^\infty \mathfrak{q}^k\frac{Z^{(k+)}(z)\pm Z^{(k-)}(z)}{2},
\end{equation}
where the sum corresponds to $\theta=0$ and the difference corresponds to $\theta=\pi$.

Writing an instanton number $k=2\ell+\xi$, the contributions can be evaluated using JK residues
\begin{equation}
    Z^{(k\pm)}(z)=\frac{1}{|W(\OO(k)^\pm)|}\oint_{\textrm{JK}}\prod_{j=1}^{\ell(-1)}\frac{\phi_j}{2\pi i} Z^{k\pm}_{\textrm{vec}} Z^{k\pm}_{\textrm{def}}(z),
\end{equation}
Note that the subscripts here express the contribution from open strings between the corresponding D-branes.
 The concrete integral expression of the vector multiplet contribution $Z^{k\pm}_{\textrm{vec}}$ for the 5d Sp($N$) vector multiplet are given as follows \cite{Kim:2012gu}:
\begin{align}\label{Sp-contour}
  \hspace{-0.5cm}Z_{\textrm{vec}}^{k=2\ell+\xi,+}=&
	   \left(\frac{1}{\llbracket \epsilon_{1,2}\rrbracket \, \prod_{\alpha=1}^{N} \llbracket \epsilon_+\pm \fraka_\alpha  \rrbracket} \cdot \prod_{i=1}^{\ell} \frac{ \llbracket \pm \phi_i\rrbracket\llbracket 2\epsilon_+\pm 2\phi_i\rrbracket} {\llbracket \pm \phi_i +\epsilon_{1,2}\rrbracket}\right)^\xi\\
&\cdot \prod_{i=1}^{\ell} \frac{\llbracket 2\epsilon_+\rrbracket }{ \llbracket \epsilon_{1,2}\rrbracket  \llbracket \epsilon_{1,2}\pm 2\phi_i\rrbracket \, \prod_{\alpha=1}^{N} \llbracket \epsilon_+\pm \phi_i \pm \fraka_\alpha \rrbracket }
   \prod_{i<j}^{\ell} \frac{\llbracket  \pm \phi_i \pm \phi_j\rrbracket \llbracket 2\epsilon_+\pm \phi_i \pm \phi_j\rrbracket}{\llbracket \epsilon_{1,2}\pm \phi_i \pm \phi_j\rrbracket}\cr
    \hspace{-0.5cm}Z_{\textrm{vec}}^{k = 2\ell,-}=& \frac{\cosh{\epsilon_+}}{\llbracket \epsilon_{1,2}\rrbracket \,\llbracket 2\epsilon_{1,2}\rrbracket \, \prod_{\alpha=1}^{N} \llbracket \pm 2\fraka_\alpha  + 2\epsilon_+\rrbracket} \cdot \prod_{i=1}^{\ell-1} \frac{ \llbracket \pm 2\phi_i\rrbracket  \llbracket 4\epsilon_+\pm 2\phi_i\rrbracket } {\llbracket 2\epsilon_{1,2}\pm 2\phi_i\rrbracket } \cr
	    &\cdot \prod_{i=1}^{\ell-1} \frac{\llbracket 2\epsilon_+\rrbracket }{ \llbracket \epsilon_{1,2} \rrbracket  \llbracket \epsilon_{1,2}\pm 2\phi_i \rrbracket \, \prod_{\alpha=1}^{N} \llbracket \epsilon_+\pm \phi_i \pm \fraka_\alpha  \rrbracket }
		\prod_{i<j}^{\ell-1} \frac{\llbracket 2\epsilon_{+}\pm\phi_i\pm\phi_j\rrbracket\,
\llbracket \pm\phi_i\pm\phi_j\rrbracket }{\llbracket \epsilon_{1,2}\pm\phi_i\pm\phi_j\rrbracket}  \cr
\hspace{-0.5cm}Z_{\textrm{vec}}^{k=2\ell+1,-}=& \frac{1}{\llbracket \epsilon_{1,2}\rrbracket \, \prod\limits_{\alpha=1}^{N} 2\cosh{ \frac{\epsilon_+\pm  \fraka_\alpha}{2}}} \cdot \prod_{i=1}^{\ell} \frac{ 2\cosh{\tfrac{\pm \phi_i}{2}} 2\cosh{ \frac{ 2\epsilon_+\pm \phi_i }{2}}} {2\cosh{ \frac{\e_{1,2}\pm \phi_i}{2}}}\cr
	&\cdot \prod_{i=1}^{\ell} \frac{\llbracket 2\epsilon_+\rrbracket }{ \llbracket \epsilon_{1,2}\rrbracket  \llbracket \epsilon_{1,2}\pm 2\phi_i\rrbracket \, \prod_{\alpha=1}^{N} \llbracket \epsilon_+\pm \phi_i \pm \fraka_\alpha \rrbracket }
   \prod_{i<j}^{\ell} \frac{\llbracket  \pm \phi_i \pm \phi_j\rrbracket \llbracket 2\epsilon_+\pm \phi_i \pm \phi_j\rrbracket}{\llbracket \epsilon_{1,2}\pm \phi_i \pm \phi_j\rrbracket}~.\nonumber
	\end{align}
Taking into consideration the Cartan subalgebras of $\OO(k)^\pm$ \cite{Kim:2012gu}, the defect contributions can be obtained from Figure \ref{tab:field-content}:
\begin{align}\label{C-defect}
    Z^{k=2\ell+\xi,+}_{\textrm{def}}(z)=&\prod_{\alpha=1}^N\llbracket \zeta\pm \mathfrak{a}_\alpha\rrbracket\lt(\frac{\llbracket \epsilon_-\pm\zeta\rrbracket}{\llbracket \epsilon_+\pm\zeta\rrbracket}\rt)^\xi\prod_{i=1}^\ell \cS(\zeta\pm\phi_i),\cr
Z^{k=2\ell+1,-}_{\textrm{def}}(z)=&\prod_{\alpha=1}^N\llbracket \zeta\pm \mathfrak{a}_\alpha\rrbracket\frac{\cosh(\frac{\epsilon_-\pm\zeta}2)}{\cosh(\frac{\epsilon_+\pm\zeta}2)}\prod_{i=1}^\ell \cS(\zeta\pm\phi_i),\cr
Z^{k=2\ell,-}_{\textrm{def}}(z)=&\prod_{\alpha=1}^N\llbracket \zeta\pm \mathfrak{a}_\alpha\rrbracket\frac{\llbracket 2\epsilon_-\pm 2\zeta\rrbracket}{\llbracket 2\epsilon_+\pm 2\zeta\rrbracket}\prod_{i=1}^{\ell-1}\cS(\zeta\pm\phi_i)~,
\end{align}
which are the same as those in \cite{Haouzi:2020yxy}. Note that the orders of the Weyl groups are given by
\begin{align}
    |W(\OO(2\ell)^+)|=&\frac{1}{2^{\ell-1}\ell!},\quad |W(\OO(2\ell{+}1)^+)|=\frac{1}{2^{\ell}\ell!},\cr |W(\OO(2\ell)^-)|=&\frac{1}{2^{\ell-1}(\ell{-}1)!},\quad |W(\OO(2\ell{+}1)^-)|=\frac{1}{2^{\ell}\ell!}.
\end{align}

\subsection{qq-characters in the unrefined limit}

As presented in \cite{Haouzi:2020yxy}, for a generic $\Omega$-background $(\epsilon_1,\epsilon_2)$, the qq-characters for $BCD$-type gauge theories appear to be ``infinite'', that is, they cannot be expressed as the expectation value of a finite number of $Y$-operators (at least in an obvious way). This originates from the fact that JK poles are rather complicated (and yet to be understood) at the refined level.  Nevertheless, the lesson we learned in \cite{Nawata:2021dlk} is that non-trivial JK poles for $BCD$-type gauge theories are classified by a set of Young diagrams in the unrefined limit  $\e_1=-\e_2=\hbar$. Therefore, in this subsection, we shall show that the qq-characters for $BCD$-type gauge theories take on finite forms in the unrefined limit, thus preserving their algebraic meaning (see Appendix \ref{a:qq} for further details).

\subsubsection*{Type $BD$}

As analyzed in \cite{Nawata:2021dlk}, non-trivial JK poles of SO($n$) pure Yang-Mills theory are classified by $N$-tuples of Young diagrams where we write $n=2N+\xi$ ($n\equiv\xi \bmod 2$). More explicitly, the location of a pole is indeed the unrefined limit of \eqref{pole-location}:
\begin{equation}\label{pole-unrefine}
\phi_{x}=\mathfrak{a}_\alpha+(i-j)\hbar~,\qquad x=(i,j)\in\lambda^{(\alpha)}~.
\end{equation}
where $|\vec{\lambda}|=k$. Therefore, the unrefined instanton partition function can be written as
\begin{equation}\label{Z-SO}
    Z^{\frakso(n)}_{\textrm{inst}}=\sum_{\vec{\lambda}}\mathfrak{q}^{|\vec{\lambda}|} Z^{\frakso(n)}_{\vec{\lambda}},
\end{equation}
where explicit forms of $Z^{\frakso(n)}_{\vec{\lambda}}$ are given in \eqref{Z-SOeven-r} and \eqref{Z-SOodd-r}.
Even if we insert a co-dimension four defect, the pole structure stays intact. Hence, we write the set of poles as $\ket{\vec{A},\vec{\lambda}}$, and the expectation value of the defect at $\ket{\vec{A},\vec{\lambda}}$ is
\begin{equation}
    \cY_{\vec{\lambda}}^{BD}(z) = \llbracket \zeta\rrbracket^\xi \prod_{\alpha=1}^N\llbracket \zeta\pm \mathfrak{a}_\alpha\rrbracket\prod_{x\in\vec{\lambda}}\cS(\phi_x\pm\zeta),\label{def-YBD}
\end{equation}
Then, the total defect partition function receives the two contributions: the one comes from the poles $\ket{\vec{A},\vec{\lambda}}$ while the other terms from $\phi_x=\pm\zeta$ and $\ket{\vec{A},\vec{\lambda'}}$ with $|\vec{\lambda}^{\prime}|=|\vec{\lambda}|-1$. Thus, the qq-character takes the form as
\begin{equation}
    \langle\chi^{BD}(z)\rangle=Z_{\textrm{defect}}^{\mathfrak{so}(n)}(z)=\langle Y^{BD}(z)\rangle +\lt\langle\frac{c_{BD}\mathfrak{q}}{Y^{BD}(z)}\rt\rangle,\label{BD-qq}
\end{equation}
where
\begin{equation}
c_{BD}=\llbracket 2\zeta\pm\hbar\rrbracket\llbracket 2\zeta\rrbracket^2~.
\end{equation}
Here, the expectation value of the $Y$-operator can be evaluated in a similar way to \eqref{exp-YA}, namely by sandwiching the corresponding Gaiotto state.

In Appendix \ref{a:id}, we provide a proof of the pole cancellations at $\zeta=\pm\phi_x$ for $x\in \mathfrak{A}(\vec{\lambda})\bigcup\mathfrak{R}(\vec{\lambda})$ with the recursive relations presented there. Given the fact that all JK poles can be classified by Young diagrams, it follows directly that the qq-character \eqref{BD-qq} is a Laurent polynomial of degree $n$ in $z$. Alternatively, the astute reader can independently verify this property by explicitly evaluating the defect partition function.

Thanks to the polynomial nature of $\langle\chi^{BD}(z)\rangle$,  \eqref{BD-qq} can be interpreted as the quantization (by $q=e^{-\hbar}$) of the Seiberg-Witten curve for the $\SO(n)$ pure Yang-Mills theory \cite{Argyres:1995fw,Danielsson:1995is,DHoker:1996kdj,Brandhuber:1997cc,Landsteiner:1997vd} by appropriately rescaling $\frakq$ and $z$.

\subsubsection*{Type $C$}
Now let us move on to the $\Sp(N)$ gauge group. It was observed in \cite{Nawata:2021dlk} that non-trivial JK poles for the  $\Sp(N)$ pure Yang-Mills theory are classified by $(N+4)$-tuples of Young diagrams where there are four additional effective Coulomb branch parameters $\mathfrak{a}_{N+j}$ ($j=1,\ldots,4$) depending on the sectors:
\begin{equation}\label{effective-Coulomb}
    \mathfrak{a}_{N+j}=\begin{cases}
    \frac{\hbar}{2}(+\pi i), \  0(+\pi i)   &(\textrm{even},+)\ \textrm{sector}\\
        \frac{\hbar}{2}(+\pi i), \ \hbar, \ \pi i&(\textrm{odd},+)\ \textrm{sector}\\
            \frac{\hbar}{2}(+\pi i), \  \hbar(+\pi i)   &(\textrm{even},-)\ \textrm{sector}\\
        \frac{\hbar}{2}(+\pi i), \ 0, \ \hbar+\pi i&(\textrm{odd},-)\ \textrm{sector}
    \end{cases}
\end{equation}
where we use the notation $(\textrm{even/odd},\pm)$ to label the $\OO(k)^\pm$ with even/odd number $k$ of instantons. Moreover, the unrefined instanton partition function involves non-trivial multiplicity coefficients for these effective Coulomb branch parameters, and it  schematically takes the following form:
\begin{equation}\label{Z-Sp}
Z^{\fraksp(N),\pm}_{\textrm{inst}}=\sum_{\vec{\lambda}}\mathfrak{q}^{|\vec{\lambda}|} C_{\vec{A},\vec{\lambda}}^{(\pm)}Z^{\frakso(2N+8)}_{\vec{\lambda}}~.
\end{equation}
We refer the reader to \cite{Nawata:2021dlk} for the explicit form of the multiplicity coefficients $C_{\vec{A},\vec{\lambda}}^{(\pm)}$.

Incorporating the effective Coulomb branch parameters, the locations of poles are as in \eqref{pole-unrefine}.
In 5d, the defect contributions \eqref{C-defect} also depend on the sectors. Consequently, we can evaluate the expectation value of the defect only in each sector:
\begin{align}\label{Y-C}
  \cY_{\vec{\lambda}}^{C}(z) =&\prod_{\alpha=1}^N\llbracket \zeta\pm \mathfrak{a}_\alpha\rrbracket
\frac{\llbracket \epsilon_-\pm\zeta\rrbracket}{\llbracket \epsilon_+\pm\zeta\rrbracket} \prod_{x\in \vec{\lambda}}\cS(\zeta\pm\phi_{x}) ~, &(\textrm{odd},+)\ \textrm{sector}\cr
\cY_{\vec{\lambda}}^{C}(z) =&\prod_{\alpha=1}^N\llbracket \zeta\pm \mathfrak{a}_\alpha\rrbracket\prod_{x\in \vec{\lambda}}\cS(\zeta\pm\phi_{x}),&(\textrm{even},+)\ \textrm{sector}\cr
\cY_{\vec{\lambda}}^{C}(z) =&\prod_{\alpha=1}^N\llbracket \zeta\pm \mathfrak{a}_\alpha\rrbracket\frac{\cosh(\frac{\epsilon_-\pm\zeta}2)}{\cosh(\frac{\epsilon_+\pm\zeta}2)}\prod_{x\in \vec{\lambda}} \cS(\zeta\pm\phi_{x}), &(\textrm{odd},-)\ \textrm{sector}\cr
 \cY_{\vec{\lambda}}^{C}(z) =&\prod_{\alpha=1}^N\llbracket \zeta\pm \mathfrak{a}_\alpha\rrbracket\frac{\llbracket 2\epsilon_-\pm 2\zeta\rrbracket}{\llbracket 2\epsilon_+\pm 2\zeta\rrbracket}\prod_{x\in \vec{\lambda}} \cS(\zeta\pm\phi_{x}), &(\textrm{even},-)\ \textrm{sector}
\end{align}
Plugging these into the summation \eqref{Z-Sp}, one can evaluate the expectation value of the $Y$-operator in each sector.
Using these expressions, we can represent the defect partition function in terms of the $Y$-operator, and it is remarkably independent of a choice of a sector (even/odd,$\pm$)
\begin{equation}\label{C-qq-pre}
    Z^{(k,\pm)}(z)=\lt\langle Y^C\rt\rangle+\frac{\mathfrak{q}^2}{\llbracket2\zeta\rrbracket^2\llbracket 2\zeta\pm \hbar\rrbracket}\lt\langle\frac{1}{Y^C}\rt\rangle~.
\end{equation}
Finally, depending on the $\theta$-angle, we can obtain the defect partition function  via \eqref{C-defect-PF}.

However, unlike in the case of $ABD$-type gauge groups, due to the additional poles \eqref{effective-Coulomb}, one cannot repeat the argument of pole-cancellation, in particular, we can see that the defect partition function of Sp($N$) theory has explicit poles at $z=\pm 1$.  To address this issue, we need to subtract a spurious contribution $Z_{\textrm{extra}}$ from the defect partition function to define a regularized qq-character. The form of $Z_{\textrm{extra}}$ is given by:
\be
Z_{\textrm{extra}}^{\mathfrak{sp}(N)_\theta}(z)=\begin{cases}
\mathfrak{q} \frac{z(1+z^2)}{(1-z^2)^2}  Z_{\textrm{inst}} ^{\mathfrak{sp}(N)_\theta}     & \textrm{for} \ N -\frac{\theta}{\pi}\equiv 1 \mod 2 \cr
    \mathfrak{q}\frac{2z^2}{(1-z^2)^2} Z_{\textrm{inst}}^{\mathfrak{sp}(N)_\theta}        & \textrm{for} \  N-\frac{\theta}{\pi} \equiv 0 \mod 2
\end{cases},\label{extra-Z}
\ee
where $Z_{\textrm{inst}}$ is the instanton partition function without the defect. The regularized qq-character is then defined as:
\begin{equation}
\lt\langle\chi^{\mathfrak{sp}(N)_\theta}(z)\rt\rangle=Z_{\textrm{defect}}^{\mathfrak{sp}(N)_\theta}(z) -Z_{\textrm{extra}}^{\mathfrak{sp}(N)_\theta}(z),\label{reg-qq}
\end{equation}
which becomes a Laurent polynomial in $z$. We have verified this property for Sp(1), Sp(2), and Sp(3) theories up to six instantons using \textit{Mathematica}.

The advantage of the universal form of the right-hand side of \eqref{C-qq-pre} is now clear. By recognizing that the spurious contribution is proportional to $Z_{\textrm{inst}} ^{\mathfrak{sp}(N)_\theta}=\langle1\rangle$, we can express \eqref{C-qq-pre} as
\be \label{C-qq}
\lt\langle Y^C\rt\rangle+f(N,\theta)\frac{\frakq}{\llbracket2\zeta\rrbracket^2} \lt\langle 1\rt\rangle+\frac{\mathfrak{q}^2}{\llbracket2\zeta\rrbracket^2\llbracket 2\zeta\pm \hbar\rrbracket}\lt\langle\frac{1}{Y^C}\rt\rangle = \lt\langle\chi^{\mathfrak{sp}(N)_\theta}(z)\rt\rangle
\ee 
where 
\be 
f(N,\theta)=\begin{cases}
2\cosh(\zeta)  & \textrm{for} \ N -\frac{\theta}{\pi}\equiv 1 \mod 2 \cr
2     & \textrm{for} \  N-\frac{\theta}{\pi} \equiv 0 \mod 2
\end{cases}~.\label{f}
\ee 
In conclusion, the form of the qq-character becomes apparent only at the unrefined limit for the $\Sp(N)$ pure Yang-Mills theory.
Also, by appropriately rescaling $\frakq$ and $z$, it can be interpreted as the quantization of the Seiberg-Witten curve \cite{Brandhuber:1997cc,Landsteiner:1997vd,Li:2021rqr}. The presence of the ``spurious'' contribution is ``essential'' for obtaining the correct form as the quantization of the Seiberg-Witten curve, and it provides the explicit dependence on the $\theta$-angle in 5d. 
Furthermore, as both cases in \eqref{f} approach $2$ in the 4d limit, we can conclude that the Seiberg-Witten curve becomes insensitive to the $\theta$-angle in the 4d limit.

\paragraph{Remark} In \eqref{C-qq}, there are apparent poles at $\zeta=\pm\phi_x$ for $x\in \mathfrak{A}(\vec{\lambda})\bigcup\mathfrak{R}(\vec{\lambda})$ and also at special values as $\zeta=0,\pm\frac{1}{2},\pm\hbar$, coming from four additional Young diagrams used to describe the Sp($N$) theories. As in Appendix \ref{a:id}, it is straightforward to show the pole cancellation at a generic pole like the case of SO($n$). However, the cancellations at these special poles are highly non-trivial and we verify the cancellations only through the order-by-order instanton expansion up to 6-instanton.

\section{Lie-algebraic relations among qq-characters}\label{s:group}
In \S\ref{sec:qq}, it was discussed that a qq-character can be interpreted as a partition function involving co-dimension four defects. Specifically, the D4-D4' strings depicted in Figure \ref{tab:field-content} introduce fermionic degrees of freedom \cite{Tong:2014yna}. When a 5d gauge theory is coupled to these one-dimensional fermionic degrees of freedom, it gives rise to a half-BPS Wilson loop. Moreover, the path integral involving this fermionic Fock space serves as a generating function for half-BPS Wilson loops in antisymmetric tensor representations \cite{Gomis:2006sb, Tong:2014cha, Kim:2016qqs}. Therefore, the qq-character can be understood as 
\begin{equation}\label{qq-Wilson}
\chi^\frakg(z)=z^{-\frac{\dim \square}{2}}\sum_{k=0}^{\dim \square}(-z)^{k}{\cal W}_{\wedge^k},
\end{equation}
where $\bigwedge^k$ denotes the $k$-th anti-symmetric tensor product of the fundamental representation $\square$ of the underlying gauge algebra, and $\dim \square$ is the dimension of the fundamental representation.
At the zero-instanton level, a Wilson loop expectation value is simply the character of the corresponding representation. While it receives all the instanton corrections, the isomorphism of representations of gauge algebras leads to the agreement of the Wilson loop expectation values at all the instanton sectors. This section investigates this aspect in the qq-characters as viewed from the perspective of Wilson loops.
The results in this section support the physical interpretation of a qq-character as a generating function of Wilson loop expectation values, shedding light on the relationships between qq-characters of different gauge algebras.
However, as we will see below, it is important to note that even if two gauge algebras are isomorphic, their qq-characters may not necessarily coincide entirely due to their dependence on representations.

\subsection{\texorpdfstring{$\mathfrak{sp}(1)$ vs $\mathfrak{su}(2)$}{sp1 vs su2}}\label{sec:sp1-su2}

Let us first consider the isomorphism $\mathfrak{sp}(1)\cong\mathfrak{su}(2)$
 of the Lie algebras. Since their fundamental representations agree, we expect the match of both the qq-characters. However, there are two subtleties to consider. The first subtlety is the choice ($\theta=0,\pi$) of the $\theta$-angle as seen in \S\ref{sec:qq}. In this isomorphism of the gauge algebras, it corresponds to a choice ($\kappa=0,1$) of the 5d Chern-Simons level $\kappa$ of the $\SU(2)$ theory \cite{Haouzi:2020zls}.  The second subtlety is that the naive $\Sp(1)$ defect partition function is not a Laurent polynomial of $z$, and we need to remove the spurious contribution $Z_{\textrm{extra}}$ given in \eqref{extra-Z} in order to compare the qq-character of $\SU(2)$.

First, let us consider the case $\kappa=0$ (or equivalently $\theta=0$). The qq-character of $\SU(2)_{\kappa=0}$ is well-known \cite{BPS/CFT}:
\begin{align}\label{su2-qq}
   \langle\chi^{\mathfrak{su}(2)_{\kappa=0}}(z)\rangle=& \frac{(z-A)(z-A^{-1})}{z}\cr
& + \frakq\frac{A(2qA(1+z^2)-z(1+q^2)(1+A^2)}{z(1-q)^2(1-A^2)^2}+\cO(\frakq^2)~,
\end{align}
At zero-instanton level, it can be expanded as $z^{-1}-(A+A^{-1})+z$, which is a generating function of the $\fraksu(2)$ characters of $(\emptyset,\square,\yng(1,1))$.

On the other hand, the defect partition function of $\Sp(1)_{\theta=0}$ is
\begin{align}\label{Sp1-defect}
Z_{\textrm{defect}}^{\mathfrak{sp}(1)_{\theta=0}}(z)=&\frac{(z-A)(z-A^{-1})}{z}\\
& +\frakq\frac{(z-A)(1-z A)}{z(1-z^2)^2(1-q)^2(1-A^2)^2}\cr
    &\times\lt(z(1+z^2)(1-q)^2(1+A^2)-2(q(1+z^4)-2z^2(1-q+q^2))A\rt)\cr &+\cO(\frakq^2)\nonumber
\end{align}
The  zero-instanton part comes from $Z_{\textrm{D4/D4'}}(z)$, which is equal to that of \eqref{su2-qq}. However, the one-instanton part is no longer a Laurent polynomial of $z$ due to the presence of $(1-z^2)^2$ in the denominator.
The difference between \eqref{su2-qq} and \eqref{Sp1-defect} at one-instanton is given by
\begin{equation}
    -\frac{z(1+z^2)}{(1-z^2)^2}~.
\end{equation}
Even at higher-instanton, it can be verified that
 the regularized qq-character defined in \eqref{reg-qq} is a Laurent polynomial of $z$, and moreover the following identity holds
\be
\langle\chi^{\mathfrak{su}(2)_{\kappa=0}}(z)\rangle=\langle\chi^{\mathfrak{sp}(1)_{\theta=0}}(z)\rangle=Z_{\textrm{defect}}^{\mathfrak{sp}(1)_{\theta=0}}(z)-\mathfrak{q}\frac{z\left(1+z^2\right)}{(1-z^2)^2}Z_{\textrm{inst}}^{\mathfrak{sp}(1)_{\theta=0}}~.
\ee
This identity has been checked up to 6-instanton.

Next, we consider the case $\kappa=1$ (or equivalently $\theta=\pi$). Using the procedure outlined in \eqref{int-Zvec}, we include the 5d Chern-Simons level $\kappa=1$ and perform a JK residue integral to obtain the qq-character of $\SU(2)_{\kappa=1}$:
\begin{align}\label{su2-qq-2}
   \langle\chi^{\mathfrak{su}(2)_{\kappa=1}}(z)\rangle=& \frac{(z-A)(z-A^{-1})}{z}\cr
& + \frakq\frac{(1+z^2) q(A+A^3)-z(q+2(1-q+q^2) A^2+q A^4)}{z(1-q)^2\left(1-A^2\right)^2}\cr &+\cO(\frakq^2)~,
\end{align}

On the other hand, the defect partition function of $\Sp(1)_{\theta=\pi}$ is given by
\begin{align}\label{Sp1-defect-2}
  &Z_{\textrm{defect}}^{\mathfrak{sp}(1)_{\theta=\pi}}(z)=\cr &\frac{(z-A)(z-A^{-1})}{z}\\
  &+\frakq\frac{(z-A)(1-A z)}{ z(1-z^2)^2(1-q)^2(1-A^2)^2}\cr &\times {(2 z(A{+}z)(1{+}A z){-}(1{+}z^2)(4 A z{+}(1{+}A^2)(1{+}z^2)) q{+}2 z(A{+}z)(1{+}A z) q^2)}\cr &+\cO(\frakq^2)\nonumber
\end{align}
While the zero-instanton part remains unchanged, the higher-instanton part will differ from the previous case. Now, the difference between \eqref{su2-qq-2} and \eqref{Sp1-defect-2} at one-instanton is given by
$$-\frac{2z^2}{(1-z^2)^2}~.$$
Therefore, as in  \eqref{reg-qq}, we can regularize it by a product of this factor and $\frakq Z_{\textrm{inst}}^{\mathfrak{sp}(1)_{\theta=\pi}}$ to get the
 the qq-character of $\Sp(1)_{\theta=\pi}$, which is equal to \eqref{su2-qq-2}
\be
\langle\chi^{\mathfrak{su}(2)_{\kappa=1}}(z)\rangle=\langle\chi^{\mathfrak{sp}(1)_{\theta=\pi}}(z)\rangle=Z_{\textrm{defect}}^{\mathfrak{sp}(1)_{\theta=\pi}}(z)-\mathfrak{q}\frac{2z^2}{(1-z^2)^2}Z_{\textrm{inst}}^{\mathfrak{sp}(1)_{\theta=\pi}}~.
\ee
We checked this identity up to 6-instanton.

\subsection{\texorpdfstring{$\mathfrak{sp}(2)$ vs $\mathfrak{so}(5)$}{sp2 vs so5}}

The comparison between the qq-characters of $\Sp(2)_{\theta=0}$ and $\SO(5)$ is  more subtle. First of all, the second anti-symmetric tensor product of the fundamental representation $\boldsymbol{4}$ of $\mathfrak{sp}(2)$ is reducible:
\be\label{sp2-so5}\wedge^2 \boldsymbol{4} \cong \boldsymbol{5}\oplus \boldsymbol{1}~,\ee
where the first irreducible representation $\boldsymbol{5}$ is isomorphic to the vector representation $\boldsymbol{5}$ of $\mathfrak{so}(5)$. As in  \eqref{reg-qq}, the regularized qq-character of $\Sp(2)_{\theta=0}$
\begin{equation}
   \langle\chi^{\mathfrak{sp}(2)_{\theta=0}}(z)\rangle= Z_{\textrm{defect}}^{\mathfrak{sp}(2)_{\theta=0}}-\mathfrak{q}\frac{2z^2}{(1-z^2)^2}Z_{\textrm{inst}}^{\mathfrak{sp}(2)_{\theta=0}}~,
\end{equation}
and the Wilson loop expectation value $\cW_{\wedge^2 \boldsymbol{4}}^{\mathfrak{sp}(2)_{\theta=0}}$ is the coefficient of its $z^0$-term. On the other hand, the Wilson loop expectation value $\cW_{\boldsymbol{5}}^{\mathfrak{so}(5)}$ is the coefficient of the $z^{-\frac32}$-term of the qq-character of SO(5), and the trivial representation $\boldsymbol{1}$ in \eqref{sp2-so5} correspond to the instanton partition function without defect. Hence, the isomorphism \eqref{sp2-so5} gives rise to the identity
\begin{equation}
     \langle\chi^{\mathfrak{sp}(2)_{\theta=0}}(z)\rangle\big|_{z^0,A_1\to A_1 A_2}= \big(\langle\chi^{\mathfrak{so}(5)}(z)\rangle\big|_{z^{-\frac32}}+Z_{\textrm{inst}}^{\mathfrak{so}(5)}\big)_{A_2\to A_1A_2^2}~.
\end{equation}
We also checked this proposal up to 6-instanton.

One may wonder about the realization of $\Sp(2)_{\theta=\pi}$ in terms of $\SO(5)$. This can be achieved by adding a spinor hypermultiplet and taking a suitable limit of the spinor mass \cite[Eq.(3.25)]{Chen:2023smd}. We expect that this can be verified at the level of qq-characters by inserting a co-dimension four defect, but we relegate this issue to future work.

\subsection{\texorpdfstring{$\mathfrak{so}(4)$ vs $\mathfrak{su}(2)\oplus \mathfrak{su}(2)$}{so4 vs su2+su2}}

As noted above, we should be careful about the representations we consider respectively in $\mathfrak{so}(4)$ and $\mathfrak{su}(2)\oplus \mathfrak{su}(2)$ to extract out the corresponding coefficients that give the same ${\cal W}_{\wedge^k}$. The vector representation $\boldsymbol{4}$ of $\mathfrak{so}(4)$ is isomorphic to the product of the two fundamental representations $\boldsymbol{2}$ of $\mathfrak{su}(2)$, i.e. $\boldsymbol{4}\cong \boldsymbol{2}\otimes \boldsymbol{2}$, so it is natural to expect that
\begin{equation}
    \cW^{\mathfrak{so}(4)}_{\boldsymbol{4}}\overset{?}{=}\cW^{\mathfrak{su}(2)}_{\boldsymbol{2}}\times \cW^{\mathfrak{su}(2)}_{\boldsymbol{2}}.
\end{equation}
However, in 5d $\cN=1$ gauge theory, an extra U(1) factor needs to be included. As shown in \cite{Hayashi:2020hhb}, the partition function of pure 5d $\cN=1$ SO(4) gauge theory can be written as
\begin{equation}
    Z_{\SO(4)}(A_1,A_2;\mathfrak{q})=Z_{\SU(2)}(A^{\frac12}_1/A_2^{\frac12},\mathfrak{q})Z_{\SU(2)}(A_1^{\frac12}A_2^{\frac12},\mathfrak{q})Z_{\textrm{U}(1)}(\mathfrak{q}),\label{SO4-SU2}
\end{equation}
where the U(1) instanton partition function, expressed in terms of the plethystic exponential
\begin{equation}
    Z_{\textrm{U}(1)}(\mathfrak{q})=\textrm{P.E.}\lt(\frac{q}{(1-q)^2}\mathfrak{q}\rt),
\end{equation}
corresponds to the contribution from the parallel branes colored in red in the 5-brane web construction of SO(4) shown below:
\begin{align}
\begin{tikzpicture}
\draw [dashed] (-1,0)--(5,0);
\draw (0,0)--(1,0.5);
\draw (4,0)--(3,0.5);
\draw (1,0.5)--(3,0.5);
\draw (1,0.5)--(1.5,1);
\draw (3,0.5)--(2.5,1);
\draw (1.5,1)--(2.5,1);
\draw[ultra thick,red] (1.5,1)--(1.5,1.5);
\draw[ultra thick,red] (2.5,1)--(2.5,1.5);
\draw[<->,dotted] (0,0.05)--(0,0.45);
\node at (0,0.25) [left] {$P$};
\draw[<->,dotted] (0,0.55)--(0,0.95);
\node at (0,0.75) [left] {$Q'$};
\draw[<->,dotted] (1.6,1.1)--(2.4,1.1);
\node at (2,1.1) [above] {$Q$};
\node at (2,0) [below] {O5$^-$};
\node at (-0.5,0) [below] {O5$^+$};
\node at (4.7,0) [below] {O5$^+$};
\end{tikzpicture}
\label{o-vert-so4}
\end{align}
Therefore, the isomorphism of the representations for the Wilson loop expectation values amounts to
\begin{equation}
    \cW^{\mathfrak{so}(4)}_{\boldsymbol{4}}(A_1,A_2)=\cW^{\mathfrak{su}(2)}_{\boldsymbol{2}}(A_1^{\frac12}/A_2^{\frac12})\times \cW^{\mathfrak{su}(2)}_{\boldsymbol{2}}(A_1^{\frac12}A_2^{\frac12})\times Z_{\textrm{U}(1)}~.\label{W-SO4-SU2}
\end{equation}
From \eqref{qq-Wilson}, the Wilson loop with the vector representation of $\mathfrak{so}(4)$ is the coefficient of $z^{-1}$-term  of the qq-character $\chi^{\mathfrak{so}(4)}$
\begin{align}\label{so4}
    \cW^{\mathfrak{so}(4)}_{\boldsymbol{4}}= & -\langle\chi^{\mathfrak{so}(4)}(z)\rangle\big|_{z^{-1}}\cr
    =&\frac{(A_1+A_2)(1+A_1A_2)}{A_1A_2}+\mathfrak{q}\frac{(A_1+A_2)(1+A_1A_2)}{(1-q)^2A_1A_2(A_1-A_2)^2(1-A_1A_2)^2}\cr
    &\times \lt(A_1A_2(1+A_2^2)(1-q)^2+A_1^3A_2(1+A_2^2) (1-q)^2+A_2^2q+A_1^4A_2^2q\rt.\cr
  & \qquad \lt.+A_1^2(q+A_2^4q-4A_2^2(1-q+q^2))\rt)+\cO\lt(\mathfrak{q}^2\rt),
\end{align}
Likewise, \eqref{qq-Wilson} indicates that the Wilson loop with the fundamental representation of $\mathfrak{su}(2)$ is the coefficient of $z^{0}$-term of the qq-character \eqref{su2-qq}
\begin{equation}\label{su2}
    \cW^{\mathfrak{su}(2)}_{\boldsymbol{2}}=-\langle\chi^{\mathfrak{su}(2)}(z)\rangle\big|_{z^{0}}=A+A^{-1}+\mathfrak{q}\frac{A(1+q^2)(1+A^2)}{(1-q)^2(1-A^2)^2}+\cO\lt(\mathfrak{q}^2\rt)~.
\end{equation}
It is clear from \eqref{so4} and \eqref{su2} that the zero-instanton part is simply the character of the corresponding representation.
In addition, we have verified the proposed identity \eqref{W-SO4-SU2} using \textit{Mathematica} up to 4-instanton.

\subsection{\texorpdfstring{$\mathfrak{so}(6)$ vs $\mathfrak{su}(4)$}{so6 vs su4}}

The vector representation $\boldsymbol{6}$ of $\mathfrak{so}(6)$  is isomorphic to the rank-two antisymmetric representation ${\yng(1,1)}$ of $\mathfrak{su}(4)$. The Wilson loop expectation value $\cW^{\mathfrak{so}(6)}_{\boldsymbol{6}}$ can be read off from the coefficient of $z^{-2}$-term in $\chi^{\mathfrak{so}(6)}$. On the other hand,  \eqref{qq-Wilson}  tells us that $\cW^{\mathfrak{su}(4)}_{\yng(1,1)}$ is equal to the coefficient of $z^0$-term in $\chi^{\mathfrak{su}(4)}$. Therefore, we have the identity
\begin{equation}
 \langle  \chi^{\mathfrak{so}(6)}(z)\rangle\big|_{z^{-2}}=\langle\chi^{\mathfrak{su}(4)}(z)\rangle\big|_{z^{0}}.\label{W-SU4-SO6}
\end{equation}
where we make the following change of the Coulomb branch parameters from the orthogonal basis of  $\mathfrak{su}(4)$ to that of $\mathfrak{so}(6)$
$$    A_1\to (A_1A_2A_3)^{\frac{1}{2}},\  A_2\to (A_1A_2^{-1}A_3^{-1})^{\frac{1}{2}},\  A_3\to (A_1^{-1}A_2A_3^{-1})^{\frac{1}{2}},\  A_4\to (A_1^{-1}A_2^{-1}A_3)^{\frac{1}{2}}~.$$
It is straightforward to check the equality \eqref{W-SU4-SO6} with \textit{Mathematica}, and we checked it up to 4-instanton.

\section{Ward-identity approach to qq-character and algebraic structure}\label{s:qq-alg}

Another interesting aspect of the qq-characters comes from their connection to the algebraic structures of supersymmetric gauge theories. In the context of 5d $A$-type gauge theories, qq-characters generate the quiver $\cW$-algebra of the gauge theory \cite{Kimura:2015rgi} and can be embedded in the quantum toroidal algebra $\QTA$ of $\mathfrak{gl}_1$ \cite{Bourgine:2017jsi}, which frames the topological vertex formalism of 5d gauge theories \cite{Awata:2011ce}.  In the latter algebraic framework, there exists a purely algebraic method for deriving qq-characters through the use of a trivial identity, referred to as a ``Ward identity'', in the quantum toroidal algebra \cite{BMZ,5dBMZ}. The existence of such an algebraic structure is deeply related to the integrability nature of the gauge theory we are considering. Unfortunately, at the current stage we do not know for sure if the $BCD$-type gauge theories have integrable structures on the generic $\Omega$-background. In this section, we try to construct a candidate for the underlying algebraic structure of  $BD$-type gauge theories by employing the ``Ward identity'' approach to qq-characters in these theories inversely. It may uncover some integrability nature of $BD$-type gauge theories in the unrefined limit. 

\subsection{Quantum toroidal algebra}

The underlying algebraic structure of $A$-type gauge theories is governed by the quantum toroidal algebra $\QTA$ of $\frakgl_1$, which is defined by the Drinfeld currents $x^\pm(z)$, $\psi^\pm (z)$ with the commutation relations \cite{DI,Miki,FFJMM1}
\begin{align}
[\psi^\pm(z),\psi^\pm(w)]= & 0,\cr
\psi^+ (z) \, \psi^- (w)= &
\frac{
g \lt( \gammah      z/w \rt)
}{
g \lt( \gammah^{-1} z/w \rt)
}
\psi^- (w) \, \psi^+ (z)
\cr
\psi^\pm (z) \, x^+ (w)= & g \lt( \gammah^{\pm \frac12} z/w \rt)    x^+ (w) \, \psi^\pm (z)
\\
\psi^\pm (z) \, x^- (w)= & g \lt( \gammah^{\mp \frac12} z/w \rt)^{-1} x^-(w) \, \psi^\pm (z)
\cr
x^\pm (z) \, x^\pm (w) = & g \lt( z/w \rt)^{\pm 1} x^\pm (w) \, x^\pm (z)
\cr
\left[x^+(z), x^-(w)\right] = &
-\lt(\Res_{z\rightarrow 1}s(z)\rt)
\lt( \delta \lt( \gammah      w/z \rt)  \, \psi^+ \lt( \gammah^{  \frac12} w \rt) -
    \delta \lt( \gammah^{-1} w/z \rt)  \, \psi^- \lt( \gammah^{- \frac12} w \rt)
\rt)~.\nonumber
\end{align}
Here $\gammah$ is a central element of the algebra, $q_3:=q^{-1}_1 q^{-1}_2$, and the structure function $g(z)$ is given by
\begin{equation}
    g(z)=\frac{s(z)}{s(z^{-1})}=\frac{(1-q_1z)(1-q_2z)(1-q_3z)}{(1-q_1^{-1}z)(1-q_2^{-1}z)(1-q_3^{-1}z)},\quad s(z):=\frac{(1-zq_1)(1-zq_2)}{(1-z)(1-zq_1q_2)}~,\label{g-DIM}
\end{equation}
which satisfies a nice property,
\begin{equation}
g(z^{-1})=g(z)^{-1}.
\end{equation}
Taking the mode expansions of the Drinfeld currents as
\begin{equation}
x^\pm(z)=:\sum_{n\in\mathbb{Z}}x^\pm_nz^{-n},\quad \psi^\pm(z)=:\sum_{\pm n\geq0}\psi^\pm_nz^{-n},
\end{equation}
an important relation, the Serre relation, needs to be imposed:
\begin{equation}
\lt[x^{\pm}_n,\lt[x^{\pm}_{n-1},x^{\pm}_{n+1}\rt]\rt]=0,
\end{equation}
for $^\forall n\in\mathbb{Z}$. Then, the central elements of the algebra are
$\gammah$ and $\psi^\pm_0$, which are mapped to constant numbers in representations. It is often convenient to parameterize them with two new operators $\hat{\ell}_1$ and $\hat{\ell}_2$,
\begin{equation}
(\gammah,\psi^\pm_0)=(\gamma^{\hat{\ell}_1},\gamma^{\mp\hat{\ell}_2}),
\end{equation}
with $\gamma:=q_3^{\frac{1}{2}}$, where we used the automorphism $\psi^\pm(z)\rightarrow \a^2\psi^\pm(z)$ and $x^\pm(z)\rightarrow \a x^\pm(z)$ for $\a\in \bC^\times$.

Now, we proceed to examine the unrefined limit $(q_1q_2=q_3^{-1}= 1)$ of this algebra. Note that in this limit, the function $g(z)$ tends to $1$ so that a naive reduction would lead to a commutative algebra at a first glance. However, this is not the case. By introducing the following mode expansion
\begin{equation}\label{Ph}
    \psi^\pm(z\gammah^{\frac12})=\psi^\pm_0\exp\lt(\sum_{m>0}\kappa_m h_{\pm m}z^{\mp m}\rt),
\end{equation}
where $\kappa_m:=(1-q_1^m)(1-q_2^m)(1-q_3^m)$, the commutation relations of $\QTA$  can be re-expressed as \cite{Feigin:2015raa}
\begin{align}
    &\lt[h_m,x_n^\pm\rt]=\mp\frac{1}{m}x^\pm_{m+n}\gammah^{-(m\pm |m|)/2},\qquad \lt[h_m,h_n\rt]=-\delta_{n+m,0}\frac{\gammah^m-\gammah^{-m}}{m\kappa_m}.
\end{align}
Then, in the unrefined limit $q_3\rightarrow 1$, we obtain \cite{Harada-master,Sasa:2019rbk}
\begin{align}
    &\lt[h_m,x_n^\pm\rt]=\mp\frac{1}{m}x^\pm_{m+n},\quad \lt[h_m,h_n\rt]=-\delta_{n+m,0}\frac{\hat{\ell}_1}{m\llbracket\pm m\hbar\rrbracket }~,\cr
    &\lt[x^+_m,x^-_n\rt]=-\llbracket \pm\hbar\rrbracket \big[(m+n)\llbracket\pm(m+n)\hbar\rrbracket h_{m+n}+\delta_{n+m,0}(n\hat{\ell}_1+\hat{\ell}_2)\big],
\end{align}
with $q_1=q_2^{-1}=q=e^{-\hbar}$. Here we use the concise notation $\llbracket\pm m\hbar\rrbracket :=(1-q^{m})(1-q^{- m})$. In terms of the Drinfeld currents, the unrefined limit of $\QTA$  can be rewritten as
\be\begin{aligned}
    &[h_m,x^\pm(z)]=\mp \frac{z^m}{m}x^\pm (z)~,\cr
    &[x^+(z),x^-(w)]=-\llbracket \pm\hbar\rrbracket\big(\delta(z/w)\sum_{\substack{m\in\mathbb{Z}\\m\neq0}}m\llbracket\pm m\hbar\rrbracket h_{m}z^{-m}+\sum_n(z/w)^n(n\hat{\ell}_1+\hat{\ell}_2)\big).
    \label{alg-unref-2}
\end{aligned}\ee

A family of representations of the algebra known as the \emph{vertical representations} is of relevance to the current context. Under the vertical representations, the algebra acts on the fixed point sets $\{\ket{\vec{A},\vec{\lambda}}_{\textrm{ref}} \}$ of the equivariant actions of the instanton moduli spaces. Moreover, the Drinfeld currents $x^\pm(z)$ add/remove a box to $N$-tuples $\vec{\lambda}$ of Young diagrams. In other words, they act as instanton creation/annihilation operators. The Cartan part $\psi^\pm(z)$ of the Drinfeld currents is related to the $Y$-operator \eqref{YA}.

\paragraph{Vertical representations}

The vertical representation maps
\begin{equation}
    (\hat{\ell}_1,\hat{\ell}_2)\mapsto (0,N),
\end{equation}
for some positive integer $N$. With a suitable normalization of the basis, the Drinfeld currents act under the representation as
\begin{align}
&x^+(z)\ket{\vec{A},\vec{\lambda}}_{\textrm{ref}}= \sum_{x\in \frakA(\vec{\lambda})}\delta(z/\chi_x)\Res_{z\rightarrow \chi_x}\frac{1}{\cY^A_{\vec{\lambda}}(z)}\ket{\vec{A},\vec{\lambda}+x}_{\textrm{ref}},\cr
&x^-(z)\ket{\vec{A},\vec{\lambda}}_{\textrm{ref}}= \sum_{x\in \frakR(\vec{\lambda})}\delta(z/\chi_x)\Res_{z\rightarrow \chi_x} \cY^A_{\vec{\lambda}}(zq_3^{-1})\ket{\vec{A},\vec{\lambda}-x}_{\textrm{ref}},\cr
&\psi^\pm(z)\ket{\vec{A},\vec{\lambda}}_{\textrm{ref}}=\lt[\frac{\cY_{\vec{\lambda}}(zq_3^{-1})}{\cY_{\vec{\lambda}}(z)}\rt]_\pm\ket{\vec{A},\vec{\lambda}}_{\textrm{ref}},\label{vert-rep-3}
\end{align}
where the $\cY^A_{\vec{\lambda}}$-function is defined in \eqref{def-YA}, and $\chi_x=\exp(-\phi_x)$. Here $[...]_\pm$ means that the expression inside the bracket is to be expanded in terms of $z^\mp$. Namely, it should be expanded in powers of $z^{-1}$ for $\psi^+$ while in powers of $z$ for $\psi^-$.

In the unrefined limit, we use the identity
\begin{equation}
    (1-q^n)^{\pm 1}=\exp\lt(\pm \log(1-q^n)\rt)=\exp\lt(\mp \sum_{m=1}^\infty \frac{1}{m}q^{nm}\rt),
\end{equation}
and \eqref{y-x-1}, \eqref{y-x-2} to obtain\footnote{We changed the normalization condition of the basis here, $\ket{\vec{A},\vec{\lambda}}=n(\vec{\lambda})\ket{\vec{A},\vec{\lambda}}_{\textrm{ref}}$. We define $n(\vec{\lambda})$ via the following recursive relation with $n(\vec{\emptyset})=1$,
\begin{equation}
    \llbracket \hbar\rrbracket\frac{n(\vec{\lambda}+x)}{n(\vec{\lambda})}=\frac{\prod_{y\in \frakR(\vec{\vec{\lambda}})}\llbracket \phi_x-\phi_y\rrbracket }{\prod_{\substack{y\in \frakA(\vec{\vec{\lambda}})\\y\neq x}}\llbracket \phi_x-\phi_y\rrbracket },
\end{equation}
and then
\begin{equation}
    \llbracket -\hbar\rrbracket\frac{n(\vec{\lambda}-x)}{n(\vec{\lambda})}=\frac{\prod_{y\in \frakA(\vec{\lambda})}\llbracket \phi_x-\phi_y\rrbracket }{\prod_{\substack{y\in \frakR(\vec{\lambda})\\y\neq x}}\llbracket \phi_x-\phi_y\rrbracket }.
\end{equation}}
\begin{align}
x^+(z)\ket{\vec{A},\vec{\lambda}}=&\llbracket \hbar\rrbracket\sum_{x\in \frakA(\vec{\lambda})}\delta(z/\chi_x)\ket{\vec{A},\vec{\lambda}+x}~,\cr
x^-(z)\ket{\vec{A},\vec{\lambda}}=&-\llbracket \hbar\rrbracket\sum_{x\in \frakR(\vec{\lambda})}\delta(z/\chi_x)\ket{\vec{A},\vec{\lambda}-x}~,\cr
h_m\ket{\vec{A},\vec{\lambda}}=&\sum_{\alpha=1}^N\frac{A_\alpha^m}{m}\frac{p_m(q^{(\lambda^{(\alpha)})^{t}+\rho+1/2})}{1-q^{m}}\ket{\vec{A},\vec{\lambda}}~,\cr
h_{-m}\ket{\vec{A},\vec{\lambda}}=&-\sum_{\alpha=1}^N\frac{A_\alpha^{-m}}{m}\frac{p_{-m}(q^{(\lambda^{(\alpha)})^{t}+\rho+1/2})}{1-q^{-m}}\ket{\vec{A},\vec{\lambda}},\label{vrep-unref-A-ket}
\end{align}
where $p_m(x)=\sum_ix_i^m$ is the power sum function, $\rho=(-\frac{1}{2},-\frac{3}{2},-\frac{5}{2},\dots)$, and $m>0$ in the above equations. We will explain in Appendix \ref{a:rep} how to directly check that the above action becomes a representation of the unrefined limit \eqref{alg-unref-2} of $\QTA$.

\subsection{\texorpdfstring{$BD$}{BD}-type algebra}

In this paper, we propose an algebra associated to $BD$-type gauge theories in the unrefined limit defined with the following commutation relations.
\begin{align}
  \left[h_{m}, x^{\pm}(z)\right]=&\mp (z^m+z^{-m}) x^{\pm}(z) ~,\cr
    [x^+(z),x^-(w)]
=&-\llbracket \pm\hbar\rrbracket  \bigg[ (\delta(z/w)+\delta(zw)) \sum_{\substack{m \in \mathbb{Z}\\m\neq 0}} \llbracket\pm m\hbar\rrbracket  h_{m} z^{-m} \cr &\qquad\qquad + \sum_{m \in \mathbb{Z}} 2(z^{m}/w^m+z^{m}w^m) (m\hat \ell_1+ \hat \ell_2)\bigg],\label{BCD-alg}
\end{align}
This algebra also has two central elements, $\hat{\ell}_1$ and $\hat{\ell}_2$. A new feature of the above algebra that was absent in the $A$-type \eqref{alg-unref-2} is that there is an apparent automorphism realized by
\begin{equation}
    z\mapsto z^{-1},\quad h_m\mapsto h_{-m},\quad x^\pm(z)\mapsto x^\pm(z).
\end{equation}

The proposed algebra is indeed constructed from the recursion relation of $\SO(n)$ unrefined instanton partition function \cite{Nawata:2021dlk}. Since it is written as a summation of Young diagrams, it makes perfect sense to add/remove a box, which will yield the vertical representation of the algebra.

\paragraph{Vertical representation}

The vertical representation again parameterizes the central elements as $(\hat{\ell}_1,\hat{\ell}_2)\mapsto (0,N)$ for some positive integer $N$.
\begin{align}\label{BCD-vert-rep}
    &x^+(z)\ket{\vec{A},\vec{\lambda}}=\llbracket \hbar\rrbracket\sum_{x\in \frakA(\vec{\lambda})}\big( \delta(z/\chi_x) +\delta(z\chi_x)\big) \ket{\vec{A},\vec{\lambda}+x}, \cr
        &x^-(z)\ket{\vec{A},\vec{\lambda}}=-\llbracket \hbar\rrbracket\sum_{x\in \frakR(\vec{\lambda})}\big(\delta(z/\chi_x) +\delta(z\chi_x)\big) \ket{\vec{A},\vec{\lambda}-x}, \\
        &h_{m}|\vec{A}, \vec{\lambda}\rangle=\sum_{\alpha=1}^N\lt[{A_\alpha^{m}}\frac{p_{m}\left(q^{(\lambda^{(\alpha)})^{t}+\rho+1/2}\right)}{1-q^{m}}+{A_\alpha^{-m}} \frac{p_{-m}\left(q^{(\lambda^{(\alpha)})^{t}+\rho+1/2}\right)}{1-q^{-m}}\rt]|\vec{A}, \vec{\lambda}\rangle.\nonumber
\end{align}
We check that it satisfies the algebraic relations presented in \eqref{BCD-alg} in Appendix \ref{a:rep}.

\paragraph{Remark} We note that in this paper, we consider only the vertical representations of the $BD$-type algebra proposed here, specifically by setting $\hat{\ell}_1=0$. As a result, the commutation relation between $x^+(z)$ and $x^-(w)$
\begin{equation}
    [x^+(z),x^-(w)]\propto \delta(z/w)+\delta(zw),
\end{equation}
and in fact one cannot check the consistency of the term containing $\hat{\ell}_1$ (such as a horizontal representation) in the context of this paper.

Additionally, we mention that by multiplying factors $f_1(z)$ and $f_2(z)$, which obey the symmetry $f_{1,2}(z^{-1})=f_{1,2}(z)$, to $x^\pm(z)$, the commutation relations between $h_m$ and $x^\pm(z)$ remain unchanged, while the commutator between $x^+(z)$ and $x^-(w)$ transforms as follows:
\begin{equation}
[x^+(z),x^-(w)]\rightarrow f_1(z)f_2(z)[x^+(z),x^-(w)],
\end{equation}
This is due to the fact that the commutator for $\hat{\ell}_1=0$ only contains terms proportional to the $\delta$-function, $\delta(z/w)$. However, this ambiguity is not fixed within the current approach.

\paragraph{Remark 2} As previously noted, the proposed algebra is derived from the recursive relations of the instanton partition function. However, for the $C$-type gauge groups, the partition function involves non-trivial multiplicity coefficients as shown in \eqref{Z-Sp}, which are dependent on Young diagrams. As a result, it becomes challenging to consider the recursive relations of the partition function for $C$-type gauge group, and thus, uncovering the underlying algebraic structure is beyond the scope of this paper.

\paragraph{Remark 3} Let us comment about the potential relation with the quantum groups of affine Lie algebra. The quantum group of affine $\mathfrak{sl}_n$ (and also that of $\mathfrak{gl}_n$) is embedded in the quantum toroidal algebra of $\widehat{\mathfrak{gl}}_n$ \cite{Negut:2013cz,2009arXiv0903.0917T} as the zero modes. We notice that no $BCD$-type Lie algebraic structure can be found from the zero-mode part of the $BD$-type algebra proposed in this article. Another connection between the quantum toroidal algebra of $\widehat{\mathfrak{gl}}_1$ and the quantum group of affine $\mathfrak{sl}_2$ has been pursued at the level of representations in \cite{Bourgine:2022scz}, as it is easy to see that the structure function defined in \eqref{g-DIM} reduces in the limit $q_1\rightarrow \infty$ (with $q_2$ fixed to finite) to 
\begin{equation}
    g(z)\rightarrow q_2^{-1}\frac{1-q_2z}{1-q_2^{-1}z},
\end{equation}
which is nothing but the structure function of the quantum group of affine $\mathfrak{sl}_2$ with $q^2\equiv q_2$. Unfortunately the Hall-Littlewood limit $q_1\rightarrow \infty$ cannot be considered in the current approach, and a shifted version of the algebra also needs to be further constructed. A more relevant result is presented in \cite{Litvinov:2021phc}, where a boundary operator $K(u)$ associated to the Maulik-Okounkov $R$-matrix is constructed and the twisted Yangian defined by this boundary operator might be related to the candidate algebra proposed here. 

\subsection{``Ward identity'' approach}\label{s:Ward}

In \cite{BMZ,5dBMZ}, a so-called ``Ward identity'' approach was proposed to derive the qq-characters from the vertical representations of quantum toroidal algebra and its 4d sibling. In this section, we re-formulate the computation in the unrefined limit and relate the $BD$-type algebra proposed in this paper with the qq-characters presented in the previous section.

\subsubsection*{Type $A$}
To start with, let us first work out the action of the quantum toroidal algebra on the bra states $\bra{\vec{A},\vec{\lambda}}$ in the unrefined limit with the normalization,
\begin{equation}
    \langle \vec{A},\vec{\mu}\ket{\vec{A},\vec{\nu}}=\delta_{\vec{\mu},\vec{\nu}}.
\end{equation}
We can use the identity
\begin{equation}\label{bra-ket}
    \lt(\bra{\vec{A},\vec{\lambda}\pm x}x^\pm(z)\rt)\ket{\vec{A},\vec{\lambda}}=\bra{\vec{A},\vec{\lambda}\pm x}\lt(x^\pm(z)\ket{\vec{A},\vec{\lambda}}\rt),
\end{equation}
to obtain the conjugate of the vertical representation \eqref{vrep-unref-A-ket}:
\begin{align}
&\bra{\vec{A},\vec{\lambda}}x^+(z)=\llbracket \hbar\rrbracket\sum_{x\in \frakR(\vec{\lambda})}\bra{\vec{A},\vec{\lambda}-x}\delta(z/\chi_x)~,\cr
&\bra{\vec{A},\vec{\lambda}}x^-(z)=-\llbracket \hbar\rrbracket\sum_{x\in \frakA(\vec{\lambda})}\bra{\vec{A},\vec{\lambda}+x}\delta(z/\chi_x)~,\cr
&\bra{\vec{A},\vec{\lambda}}h_m=\bra{\vec{A},\vec{\lambda}}\sum_{\alpha=1}^N\frac{A_\alpha^m}{m}\frac{p_m(q^{-(\lambda^{(\alpha)})^{t}-\rho-\frac12})}{1-q^{-m}}~,\cr
&\bra{\vec{A},\vec{\lambda}}h_{-m}=-\bra{\vec{A},\vec{\lambda}}\sum_{\alpha=1}^N\frac{A_\alpha^{-m}}{m}\frac{p_{-m}(q^{-(\lambda^{(\alpha)})^{t}-\rho-\frac12})}{1-q^{m}}.\label{vrep-unref-A-bra}
\end{align}
The key step in deriving the qq-character from the vertical representation is the following ``Ward identity'' which is a trivial identity:
\begin{equation}
    \lt(\bra{\frakG}x^+_>(z)\rt)\ket{\frakG}=\bra{\frakG}\lt(x^+_>(z)\ket{\frakG}\rt),\label{triv-id}
\end{equation}
where we define the operators containing only non-negative modes in $x^\pm(z)$ as
\begin{equation}
    x^\pm_>(z):=\sum_{n\geq 0}x^\pm_nz^{-n}.
\end{equation}
 By definition, we see that
\begin{align}
     x^+_>(z)\ket{\vec{A},\vec{\lambda}}=\llbracket \hbar\rrbracket\sum_{x\in \frakA(\vec{\lambda})}\frac{1}{1-\chi_x/z}\ket{\vec{A},\vec{\lambda}+x}~,\cr
     \bra{\vec{A},\vec{\lambda}}x^+_>(z)=\llbracket \hbar\rrbracket\sum_{x\in \frakR(\vec{\lambda})}\bra{\vec{A},\vec{\lambda}-x}\frac{1}{1-\chi_x/z}~.
\end{align}
From \eqref{ZU-rec-2}, we have
\begin{align}
&x^+_>(z)\ket{\frakG}\\
=&\llbracket \hbar\rrbracket\sum_{\vec{\lambda}}\lt(\mathfrak{q}^{|\vec{\lambda}|}Z^{\fraksu(N)}_{\vec{\lambda}}\rt)^{\frac12}\sum_{x\in \frakA(\vec{\lambda})}\frac{1}{1-\chi_x/z}\ket{\vec{A},\vec{\lambda}+x}\cr
=&(-1)^{-\frac{N-1}{2}}\mathfrak{q}^{-\frac12}\sum_{\vec{\lambda}}\sum_{x\in \frakA(\vec{\lambda})}\frac{1}{1-\chi_x/z}\lt(\mathfrak{q}^{|\vec{\lambda}|+1}Z^{\fraksu(N)}_{\vec{\lambda}+x}\rt)^{\frac12}\lim_{\zeta\rightarrow\phi_x}\llbracket \zeta-\phi_x\rrbracket \cY^A_{\vec{\lambda}+x}(z)\ket{\vec{A},\vec{\lambda}+x}\cr
=&(-1)^{-\frac{N-1}{2}}\mathfrak{q}^{-\frac12}\sum_{\vec{\lambda}'}\sum_{x\in \frakR(\vec{\lambda}')}\frac{\chi_x^{-1}}{1-\chi_x/z}\lt(\mathfrak{q}^{|\vec{\lambda}'|}Z^{\fraksu(N)}_{\vec{\lambda}'}\rt)^{\frac12}\Res_{z\rightarrow\chi_x}\cY^A_{\vec{\lambda}'}(z)\ket{\vec{A},\vec{\lambda}'},\nonumber
\end{align}
and
\begin{align}
&\bra{\frakG}x^+_>(z)\\ =&\llbracket \hbar\rrbracket\sum_{\vec{\lambda}}\lt(\mathfrak{q}^{|\vec{\lambda}|}Z^{\fraksu(N)}_{\vec{\lambda}}\rt)^{\frac12}\sum_{x\in \frakR(\vec{\lambda})}\bra{\vec{A},\vec{\lambda}-x}\frac{1}{1-\chi_x/z}\cr
=&(-1)^{\frac{N-1}{2}}\mathfrak{q}^{\frac12}\sum_{\vec{\lambda}'}\bra{\vec{A},\vec{\lambda}'}\sum_{x\in \frakA(\vec{\lambda}')}\frac{\chi_x^{-1}}{1-\chi_x/z}\lt(\mathfrak{q}^{|\vec{\lambda}'|}Z^{\fraksu(N)}_{\vec{\lambda}'}\rt)^{\frac12}\Res_{z\rightarrow\chi_x}\frac{1}{\cY^A_{\vec{\lambda}'}(z)}.\nonumber
\end{align}
Then we derive from the trivial identity \eqref{triv-id} that
\begin{equation}
    \sum_i\frac{(\chi^+_i)^{-1}}{1-\chi^+_i/z}\Res_{z\rightarrow \chi^+_i}\lt\langle Y^A(z)\rt\rangle+\mathfrak{q}(-1)^N\sum_j\frac{(\chi^-_j)^{-1}}{1-\chi^-_j/z}\Res_{z\rightarrow \chi^-_i}\lt\langle\frac{1}{Y^A(z)}\rt\rangle=0,
\end{equation}
where the sets $\{\chi^\pm_i\}$ respectively run over all poles (except $z=0,\infty$) in $\cY^A_{\vec{\lambda}}(z)$ and $1/\cY^A_{\vec{\lambda}}(z)$ (functions obtained after $(Y^{A}(z))^{\pm 1}$ evaluated in the expectation value). This ``Ward identity'' suggests that the apparent poles (excluding $z=0,\infty$) in the qq-character are all cancelled, and the expectation value of the combination
\begin{equation}
\chi(z)=Y^A(z) + \frac{(-1)^N\mathfrak{q}}{Y^A(zq_3)}~,
\end{equation}
which can only have poles at $z=0$ and $z=\infty$.  A rational function that only has poles at $z=0,\infty$ on the entire complex plane must be a Laurent polynomial. From the asymptotic behavior \eqref{asym-qq-A}, it follows that up to an overall factor $z^{-\frac{N}{2}}$, the expectation value of the qq-character is a polynomial of degree $N$. For $A$-type gauge groups, similar arguments apply to all qq-characters associated with higher-weight representations and higher-rank quivers. By constructing special combinations that cancel the apparent poles, the expressions of qq-characters (in terms of $Y$-operators) can be derived completely from the underlying algebraic structure of 5d gauge theories.

\subsubsection*{Type $BD$}
In the case of $D$-type gauge theories, i.e. gauge theories with SO($2N$) gauge group, it follows from \eqref{bra-ket} that 
the bra expression of the vertical representation is
\begin{align}
    &\bra{\vec{A},\vec{\lambda}}x^+(z)=\llbracket \hbar\rrbracket\sum_{x\in \frakR(\vec{\lambda})}\bra{\vec{A},\vec{\lambda}-x}\big(\delta(z/\chi_x) + \delta(z\chi_x)\big) , \cr
        &\bra{\vec{A},\vec{\lambda}}x^-(z)=-\llbracket \hbar\rrbracket\sum_{x\in \frakA(\vec{\lambda})}\bra{\vec{A},\vec{\lambda}+x}\big(\delta(z/\chi_x) +\delta(z\chi_x)\big) , \cr
        &\bra{\vec{A}, \vec{\lambda}}h_{m}=\bra{\vec{A}, \vec{\lambda}}\sum_{\alpha=1}^N\lt[{A_\alpha^{m}}\frac{p_{m}\left(q^{(\lambda^{(\alpha)})^{t}+\rho+1/2}\right)}{1-q^{m}}+{A_\alpha^{-m}} \frac{p_{-m}\left(q^{(\lambda^{(\alpha)})^{t}+\rho+1/2}\right)}{1-q^{-m}}\rt].\cr
\end{align}
Now we split the operators $x^\pm(z)$ in a novel way into $x^{\pm}_>(z)$ and $x^{\pm}_<(z)$ in the vertical representation \eqref{BCD-vert-rep} of the $BD$-type algebra as
\begin{align}
    x^{+}_>(z)\ket{\vec{A},\vec{\lambda}}=&(1-q)\sum_{x\in \frakA(\vec{\lambda})}\bigg( \frac{1}{1-\chi_x/z}+\frac{1}{1-z\chi_x}\bigg) \ket{\vec{A},\vec{\lambda}+x},\\
    x^{+}_<(z)\ket{\vec{A},\vec{\lambda}}=&(1-q)\sum_{x\in \frakA(\vec{\lambda})}\bigg( \frac{1}{1-z/\chi_x}+\frac{1}{1-z^{-1}\chi_x^{-1}}\bigg) \ket{\vec{A},\vec{\lambda}+x}.
\end{align}
Even for this algebra, we consider the insertion of $x^{+}_>(z)$ in the inner product of the Gaiotto state $\ket{\frakG}$ of $D$-type and derive the pole cancellation condition of qq-characters from the Ward identity \eqref{triv-id}.

It is then parallel to derive
\begin{align}
&x^{+}_>(z)\ket{\frakG}\\=&(1-q)\sum_{\vec{\lambda}}\lt(\mathfrak{q}^{|\vec{\lambda}|}Z^{\frakso(2N)}_{\vec{\lambda}}\rt)^{\frac12}\sum_{x\in \frakA(\vec{\lambda})}\lt(\tfrac{1}{1-\chi_x/z}+\tfrac{1}{1-z\chi_x}\rt)
\ket{\vec{A},\vec{\lambda}+x}\cr
=&\tfrac{(1-q)}{\llbracket \hbar\rrbracket}\mathfrak{q}^{-\frac12}\sum_{\vec{\lambda}}\sum_{x\in \frakA(\vec{\lambda})}\lt(\mathfrak{q}^{|\vec{\lambda}|+1}Z^{\frakso(2N)}_{\vec{\lambda}+x}\rt)^{\frac12}
\lim_{\zeta\rightarrow\phi_x}\tfrac{\llbracket \zeta-\phi_x\rrbracket \cY^D_{\vec{\lambda}+x}(z)}{\llbracket 2\zeta\pm\hbar\rrbracket }
\lt(\tfrac{1}{1-\chi_x/z}+\tfrac{1}{1-z\chi_x}\rt)\ket{\vec{A},\vec{\lambda}+x}\cr
=&\tfrac{(1-q)}{\llbracket \hbar\rrbracket}\mathfrak{q}^{-\frac12}\sum_{\vec{\lambda}'}\sum_{x\in \frakR(\vec{\lambda}')}\lt(\mathfrak{q}^{|\vec{\lambda}'|}Z^{\frakso(2N)}_{\vec{\lambda}'}\rt)^{\frac12}\cr
&\hspace{3cm}\times\lt(\tfrac{z}{z-\chi_x}\Res_{z\rightarrow\chi_x}\tfrac{z^{-1}\cY^D_{\vec{\lambda}'}(z)}{\llbracket 2\zeta\pm\hbar\rrbracket }+\tfrac{1}{z-\chi_x^{-1}}\Res_{z\rightarrow\chi^{-1}_x}\tfrac{z^{-1}\cY^D_{\vec{\lambda}'}(z)}{\llbracket 2\zeta\pm\hbar\rrbracket }\rt)\ket{\vec{A},\vec{\lambda}'},\nonumber
\end{align}
where we used the identities in \eqref{Z-rec-D}, and the fact that
\begin{equation}
    \lim_{\zeta\rightarrow \phi_x}\llbracket \zeta-\phi_x\rrbracket \cY^D_{\vec{\lambda}}(z)=-\lim_{\zeta\rightarrow -\phi_x}\llbracket \zeta+\phi_x\rrbracket \cY^D_{\vec{\lambda}}(z).
\end{equation}
Acting on the bra state, we have
\begin{align}
&\bra{\frakG}x^{+}_>(z)\\
=&\sum_{\vec{\lambda}}\sum_{x\in \frakR(\vec{\lambda})}\bra{\vec{A},\vec{\lambda}-x}(1-q)\lt(\mathfrak{q}^{|\vec{\lambda}|}Z^{\frakso(2N)}_{\vec{\lambda}}\rt)^{\frac12}\lt(\tfrac{1}{1-\chi_x/z}+\tfrac{1}{1-z\chi_x}\rt)\cr
=&-\sum_{\vec{\lambda}}\sum_{x\in \frakR(\vec{\lambda})}\bra{\vec{A},\vec{\lambda}-x}\tfrac{(1-q)}{\llbracket \hbar\rrbracket}\mathfrak{q}^{\frac12}\lt(\mathfrak{q}^{|\vec{\lambda}|-1}Z^{\frakso(2N)}_{\vec{\lambda}-x}\rt)^{\frac12}\lt(\tfrac{1}{1-\chi_x/z}+\tfrac{1}{1-z\chi_x}\rt)\lim_{\zeta\rightarrow \phi_x}\tfrac{\llbracket \zeta-\phi_x\rrbracket \llbracket 2\zeta\rrbracket ^2}{\cY^D_{\vec{\lambda}-x}(z)}\cr
=&-\sum_{\vec{\lambda}'}\sum_{x\in \frakA(\vec{\lambda}')}\bra{\vec{A},\vec{\lambda}'}\tfrac{(1-q)}{\llbracket \hbar\rrbracket}\mathfrak{q}^{\frac12}\lt(\mathfrak{q}^{|\vec{\lambda}'|}Z^{\frakso(2N)}_{\vec{\lambda}'}\rt)^{\frac12}\cr
&\hspace{3cm}\times\lt(\tfrac{z}{z-\chi_x}\Res_{z\rightarrow\chi_x}\tfrac{z^{-1}\llbracket 2\zeta\rrbracket ^2}{\cY^D_{\vec{\lambda}'}(z)}+\tfrac{1}{z-\chi_x^{-1}}\Res_{z\rightarrow \chi_x^{-1}}\tfrac{z^{-1}\llbracket 2\zeta\rrbracket ^2}{\cY^D_{\vec{\lambda}'}(z)}\rt).\nonumber
\end{align}
Finally, we obtain the desired formula from the Ward identity \eqref{triv-id}
\begin{equation}
    \sum_i\frac{\eta_{i}^{-\sigma}}{z-\eta^\sigma_{i}}\Res_{z\rightarrow \eta^\sigma_i}\frac{\lt\langle Y^D(z)\rt\rangle}{\llbracket 2\zeta\pm\hbar\rrbracket}+\mathfrak{q}(-1)^N\sum_j\frac{\xi_{j}^{-\sigma}}{z-\xi^\sigma_j}\Res_{z\rightarrow \xi^\sigma_j}\lt\langle\frac{\llbracket 2\zeta\rrbracket ^2}{Y^D(z)}\rt\rangle=0,
\end{equation}
where $\sigma=\pm 1$, $\{\eta_i^\sigma\}=\{\chi^\sigma_{x\in \frakR(\vec{\lambda})}\}_{\vec{\lambda}}$ denotes all the poles in $\lt\langle Y^D(z)\rt\rangle$, and $\{\xi^\sigma_i\}=\{\chi^\sigma_{x\in \frakA(\vec{\lambda})}\}_{\vec{\lambda}}$ denotes all the poles in $\lt\langle\frac{1}{Y^D(z)}\rt\rangle$.  This identity shows explicitly the pole cancellation in the qq-character. The factor
\begin{equation}
c_{BD}=\llbracket 2\zeta\pm\hbar\rrbracket\llbracket 2\zeta\rrbracket^2~,
\end{equation}
appearing in \eqref{BD-qq} is successfully reproduced from the pole-cancellation condition here.  This indicates that the pole cancellation property of the qq-characters in $\SO(2N)$ theories is a direct consequence of the existence of such a $BD$-type algebra.

The derivation of the qq-character in the $B$-type case is analogous to that in the $D$-type case. By noting that the recursive relation for the partition function of SO($2N+1$) is identical to that of SO($2N$) when $\cY^D_{\vec{\lambda}}$ is replaced by $\cY^B_{\vec{\lambda}}$ (see \eqref{Z-rec-B}), it follows that the pole cancellation in qq-characters is also a direct consequence of the same $BD$-type algebra. This highlights the universality and consistency of the algebraic structure in the 5d gauge theories with $BD$-type gauge groups.

\section{Conclusion and Discussion}

In this paper, we present analytic expressions for the fundamental qq-characters of $BCD$-type gauge theories based on the pole classifications by Young diagrams in the unrefined instanton partition functions proposed in \cite{Nawata:2021dlk}. The expectation values of the qq-characters in $ABCD$-type gauge theories are shown to take the following schematic form
\begin{align}
    \lt\langle\chi(z)\rt\rangle=&\langle Y(z)\rangle+\lt\langle\frac{c(z) \frakq}{Y(z)}\rt\rangle ~, &\textrm{for} \ ABD\textrm{-type}  \ (\ref{qq-A-1},\ref{BD-qq})\cr 
  \lt\langle\chi(z)\rt\rangle=&\langle Y(z)\rangle + \widetilde c(z) \frakq\langle1\rangle+\lt\langle\frac{c(z) \frakq^2}{Y(z)}\rt\rangle ~,& \textrm{for} \  C\textrm{-type} \ \eqref{C-qq}
\end{align}
with properly defined operators $Y$. Our results demonstrate the polynomial nature of the expressions of these qq-characters and establish Lie-algebraic relations among them. Furthermore, we introduce a new algebra referred to as the $BD$-type algebra, which is closely related to the quantum toroidal algebra of $\mathfrak{gl}_1$ in the $q_1q_2\rightarrow 1$ limit. This algebra is constructed using a technique known as the ``Ward identity'', which is used to derive the qq-characters. While many of the fundamental properties and consistency of this new algebra are yet to be fully understood, it is expected to have a deep connection to the integrability aspect of SO-type gauge theories. We leave further exploration of these properties as future work.

Certainly, there are still many directions that can be explored in future research. One limitation of our work is that it only covers the unrefined limit, as the computation on the general $\Omega$-background remains a technical challenge. To overcome this limitation, one approach is to use the blowup equation, which has been generalized for gauge theories beyond $A$-type gauge groups in \cite{Kim:2019uqw}. This method can be used to solve the instanton partition functions with an integer ratio between $\epsilon_1$ and $\epsilon_2$ starting from the unrefined instanton partition function. Additionally, blowup equations with the presence of defects in 4d/5d/6d theories have been formulated in  \cite{Nekrasov:2020qcq,Jeong:2020uxz,Bonelli:2021rrg,Kim:2021gyj,Bonelli:2022iob,Chen:2021rek}. These developments open up the possibility of exploring qq-characters on more general $\Omega$-backgrounds using the unrefined expressions derived in this paper. There were also other interesting attempts to refined formulations of gauge theories with gauge groups beyond $A$-type \cite{Hayashi:2017jze,Hayashi:2021pcj} and quiver gauge theories with quiver structure beyond $A$-type \cite{Bourgine:2017rik,Kimura:2019gon,Kim:2022dbr} mainly based on the brane construction and topological vertex formalism, and it will be desired to explore how to compute qq-characters to some closed form in such methods.

We only considered the calculation of the fundamental qq-character, i.e. the defect partition function with only one defect of co-dimension four. However, it is important to note that the higher qq-characters, which involve more defects, contain additional information and their relationship with Wilson loops, particularly in $BCD$-type gauge theories, is not yet fully understood. In $A$-type gauge theories, the qq-characters, as operator-valued quantities, generate the quiver $\cW$-algebra, and they commute with the screening charge in the $\cW$-algebra. One can show that the product of topological vertices assigned to the brane web plays the role of the screening charge with the ``Ward identity'' introduced in \S\ref{s:Ward} \cite{Bourgine:2017jsi}. It is an interesting open question to explore if a similar formulation of screening charges can be done in $BCD$-type gauge theories.

We proposed a new algebra, the $BD$-type algebra, but our proposal is still in its early stages, and there are many properties of the algebra that are yet to be discovered. In particular, the extension of the algebra to the refined case is proven to be challenging,  due to the difficulty in labeling JK residues using Young diagrams in that case.  Additionally, we have only constructed vertical representations for the algebra, which are related to D5-branes in the brane web, and it is currently unclear how to construct horizontal representations with $\hat{\ell}_1\mapsto 1$. From the perspective of the brane web construction of SO-type gauge theories, it is expected that the algebraic structure for the vertical representations or the D5-branes added with an orientifold is modified while the algebra associated to NS5-branes remains unchanged.  Therefore, it is of interest to explore whether it is possible to embed the O-vertex formulated in \cite{Hayashi:2020hhb,Nawata:2021dlk}  into the coproduct of the $BD$-type algebra and the usual quantum toroidal algebra of $\mathfrak{gl}_1$ in the unrefined limit. However, it is currently challenging to envision the role that the O-vertex could play in the purely algebraic context of the $BD$-type algebra. 

The basis of the vertical representation in the quantum toroidal algebra $\QTA$ of $\mathfrak{gl}_1$ is nothing but the Macdonald symmetric polynomial \cite{FT,FFJMM1}, and in the unrefined limit, it reduces to the Schur polynomial equivalent to the character of $A$-type Lie algebras. One further obtains the generalized Macdonald polynomial introduced in \cite{AFHKSY} by taking the coproduct of the vertical representations \cite{Awata:2015hax,Fukuda-Mac}. In the current context, it is natural to expect the orthogonal and symplectic Schur polynomials, which are naturally defined from the characters of $BCD$-type Lie algebras, to play a similar role in the vertical representation constructed in \eqref{BCD-vert-rep}. The relation between such symmetric polynomials and $BCD$-type $\cW_{1+\infty}$-algebras have been explored in \cite{Bourgine-Winf}, but there seems to be a gap between the vertex-operator formulation there and our approach at the current stage.  

Our study did not include a discussion of the candidate algebra for $C$-type gauge theories, as both the instanton partition function and the qq-characters of these theories involve the multiplicity coefficients \eqref{Z-Sp} for four pieces of Young diagram summations, making it difficult to apply the ``Ward identity'' approach. However, an alternative method for computing the Sp($N$) instanton partition function using the topological vertex formalism, where the partition function does not involve the multiplicity coefficients, was proposed in \cite{Kim-Yagi}. It is worth investigating the possibility of constructing a vertex-operator formulation of this approach and studying the underlying algebraic structure.

The similarity between the integrand structure in instanton counting for supergroups and $BCD$-type groups has been noted in \cite{KP-super}, and the topological vertex formalism for supergroups has been established in \cite{Kimura:2020lmc,Noshita:2022dxv}. A potential direction for future research is to further explore this similarity and its origin in the unrefined limit.

\acknowledgments
We would like to thank Yutaka Matsuo and Go Noshita for the comments on the draft, and we are also grateful to Sung-Soo Kim, Xiaobin Li and Futoshi Yagi for discussions.  S.N. wants to express gratitude to Southeast Jiaotong University for the warm hospitality where a part of the work was carried out.
The research of S.N. is supported by National Science Foundation of China No.12050410234 and Shanghai Foreign Expert grant No. 22WZ2502100. K.Z. (Hong Zhang) thanks Shanghai city for the fund, which unfortunately cannot be disclosed here. R.Z. is supported by National Natural Science Foundation of China No. 12105198 and the High-level personnel project of Jiangsu Province (JSSCBS20210709).

\appendix

\section{Notations and conventions}\label{a:notation}

Here, we summarize the notations and conventions employed in this paper. Roughly speaking, a relationship between a 5d parameter $P_{5d}$ and a 4d parameter $p_{4d}$, which is given by
\begin{equation}
P_{5d} = \exp(-p_{4d})
\end{equation}
We define the $\Omega$-background parameters as $q_j=\exp\lt(-\e_j\rt)$, where $q_1q_2q_3=1$ is a result of the constraint $\e_1+\e_2+\e_3=0$. Additionally, we introduce the notation $\epsilon_\pm=\frac{\epsilon_1\pm\epsilon_2}{2}$, and the unrefined limit is defined as $\e_1=-\e_2=\hbar$, resulting in $2\e_+=\e_1+\e_2=0$.
The Coulomb branch parameters are represented as $A_\alpha=\exp\lt(-\mathfrak{a}_\alpha\rt)$, and the variables of a qq-character are denoted as $z_j=\exp\lt(-\zeta_j\rt)$. The poles of JK residues are represented by $\chi_x=\exp\lt(-\phi_x\rt)$, with $\phi_x =\mathfrak{a}_\alpha-\e_++(i-1) \epsilon_1+(j-1) \epsilon_2$.
Lastly, we employ the shorthand notation $\llbracket \alpha\rrbracket:=2\sinh\left(\frac{ \alpha}{2}\right)$ and $\llbracket \alpha \pm \beta \rrbracket := \llbracket \alpha + \beta \rrbracket \llbracket \alpha - \beta \rrbracket$ for convenience throughout the article. This notation has an advantage when we take the 4d limit of a partition function because we can simply take only the argument $\llbracket \a\rrbracket \to \a$ removing the double bracket, and ignore $\cosh$ terms.

\begin{figure}[ht]
    \centering
    \includegraphics[width=7.5cm]{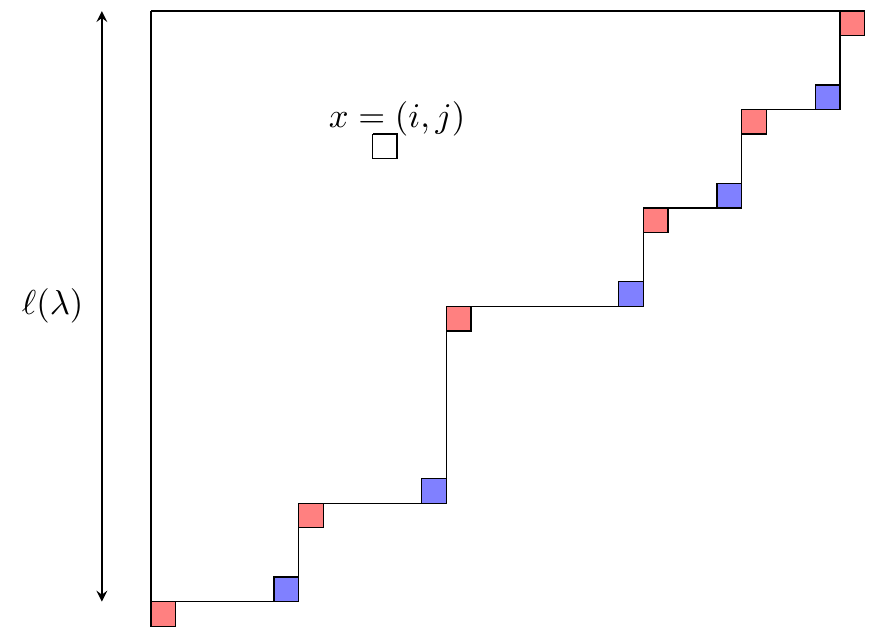}
    \caption{The red (resp. blue) boxes can be added to (resp. removed from) the Young diagram, and they form the set denoted by $\frakA(\lambda)$ (resp. $\frakR(\lambda)$).}
    \label{fig:young}
\end{figure}

In this paper, we use the following notation for Young diagrams. A Young diagram is represented by $\lambda = (\lambda_1, \lambda_2, \cdots)$, where the entries are arranged in non-decreasing order, i.e. $\lambda_1 \geq \lambda_2 \geq \cdots$. The length of a Young diagram, denoted by $\ell(\lambda)$, is defined as the number of non-zero entries in $\lambda$. The number of boxes in a Young diagram is given by $|\lambda| = \sum_{i=1}^{\ell(\lambda)} \lambda_i$, and we denote the transpose of $\lambda$ by $\lambda^t$. A set of $N$ Young diagrams is written as
$$
\vec{\lambda}=\left(\lambda^{(1)}, \ldots, \lambda^{(N)}\right)~,
$$
and the total number of boxes in this set is denoted by
$$
|\vec{\lambda}|=\sum_{\a=1}^N\left|\lambda^{(\a)}\right|~.
$$
The set of boxes in $\lambda$ that can be added or removed is represented by $\frakA({\lambda})$ or $\frakR({\lambda})$, respectively. (See Figure \ref{fig:young}.) When considering an $N$-tuple of Young diagrams, we denote the corresponding sets by $\frakA(\vec{\lambda})$ and $\frakR(\vec{\lambda})$.

We use the following notation for a plethystic exponential:
$$\text{P.E.}[f(x,y,\ldots)]=\exp \left(\sum _{k=1}^{\infty }{\frac {f(x^{k},y^k,\ldots)}{k}}\right)~,$$
which transforms a single-particle index $f$ into a multiparticle index.

\section{Finiteness of qq-characters in the unrefined limit}\label{a:qq}

In this Appendix, we present the argument that the qq-characters take the finite form in the unrefined limit whereas they involve an infinite number of terms at the refined level \cite{Haouzi:2020yxy}.

The evaluation of the partition function with or without defects follows the JK prescription. We first assign a vector $\vec{v}=\sum_{i=1}^kc_i\vec{e}_i$ to each denominator factor $\sum_ic_i\phi_i+d$, where $\vec{e}_i$ is a $k$-dimensional vector with $(\vec{e}_i)_j=\delta_{i,j}$. We also need to choose a $k$-dimensional reference vector $\Check{\eta}_k$ for each instanton number $k$, usually taken to be
\begin{equation}
    \Check{\eta}_k=(1,1+\delta_1,1+\delta_2,\dots,1+\delta_{k-1}),
\end{equation}
where $\delta_i$ are small real numbers that are not linearly dependent on integer numbers, i.e. for $^\forall I_i\in\mathbb{Z}$,
\begin{equation}
    \delta_1+\sum_{i=2}^{k-1}I_i\delta_i\neq 0.
\end{equation}
For example, for $k=2$ we can choose $\Check{\eta}_2=(1,1.001)$ and for $k=3$ we can choose $\Check{\eta}_3=(1,1.001,1.0007)$, but the choice $\Check{\eta}_3=(1,1.001,1.002)$ is not appropriate.
At the level of $k$-instanton, for a set of $k$ poles to be permitted, their corresponding vectors ${\vec{v}i}{i=1}^k$ must satisfy the following condition:
\begin{equation}
    \Check{\eta}_k=\sum_{i=1}^kc_i\vec{v}_i,\quad ^\exists c_i>0,\quad i=1,\dots,k.
\end{equation}

As an example, in the two-instanton computation of SO($2N$) theories, we have vectors $(1,0)$ attached to $\epsilon_++\phi_1\pm\mathfrak{a}_\alpha$, $(-1,0)$ attached to $\epsilon_+-\phi_1\pm\mathfrak{a}_\alpha$, $(1,1)$ corresponding to $\epsilon_++\phi_1+\phi_2\pm\epsilon_-$ and etc. In total, from the set of vectors
\begin{equation}
    \{(\pm1,0),(0,\pm1),(1,1),(1,-1),(-1,1),(-1,-1)\},
\end{equation}
the allowed combinations in the JK prescription are
\begin{align}
    \{(-1,0),(1,1)\},\ \{(-1,1),(1,1)\},\ \{(0,1),(1,-1)\},\ \{(0,1),(1,1)\},\cr
    \{(1,0),(-1,1)\},\ \{(1,0),(0,1)\}.
\end{align}
When we add the defect contribution, i.e. poles at $\epsilon_+\pm \phi_i\pm\zeta$ for $i=1,2$, the set of vectors and allowed combinations remain unchanged. However, if we pick up a pole in the defect contribution, it must be a combination containing $(\pm 1,0)$ or $(0,\pm 1)$ respectively corresponding to $\epsilon_+\pm \phi_i\pm\zeta$. If we do not pick up any new poles in the defect contribution, we get a term $\langle Y^D\rangle$, and if we pick up one pole, we get a term proportional to $\lt\langle\frac{1}{Y^D}\rt\rangle$.

To show that the qq-characters are expressed as a finite combination of expectation values of $Y$-operators in the unrefined limit, we need to rule out two possibilities. The first is that terms picking up more than one pole in the defect contribution vanish in the unrefined limit. The second is that poles of the form $\pm \zeta \pm\epsilon_+$ are allowed in the unrefined limit, but poles of the form $\pm \zeta\pm \epsilon_++i\epsilon_1+j\epsilon_2$ for $i,j\in \mathbb{Z}$ give trivial residues.

In the example of two-instanton calculation of the SO($2N$) theory shown above, we first want to exclude the possibility of the combination $\{(1,0),(0,1)\}$ selecting the poles $\epsilon_++\phi_1\pm\zeta$ and $\epsilon_++\phi_2\pm\zeta$. It is straightforward to demonstrate that these poles vanish even in the refined $\Omega$-background. This is because among the variables $\epsilon_+\pm \phi_1\pm\phi_2$, at least one must be equal to $\epsilon_+$ or $-\epsilon_+$ whenever one of the above combinations of poles is chosen.  The factor $\cS(\epsilon_+\pm \phi_1\pm\phi_2)^{-1}$ in the integrand then introduces an additional zero, eliminating any contribution from this case. This argument holds for all cases we consider in this paper and for an arbitrary number of instantons.

The second circumstance is more involved, as its contribution is non-zero in the refined case, but only vanishes in the unrefined limit. For example in the two-instanton calculation of SO($2N$) theory, picking up the pole $\phi_1=\zeta-\epsilon_+-\epsilon_1$, $\phi_2=\zeta-\epsilon_+$ corresponding to the combination $\{(0,1),(1,-1)\}$ gives non-trivial residue:
\begin{align}
  \frac{\llbracket-(\epsilon_1+\epsilon_2)\rrbracket \llbracket 2(\epsilon_1-\zeta)\rrbracket  \llbracket \epsilon_1-2\zeta\rrbracket \llbracket 2\zeta-\epsilon_2\rrbracket \llbracket 2\zeta-\epsilon_1-\epsilon_2\rrbracket \llbracket 2(-\zeta+2\epsilon_1+\epsilon_2)\rrbracket }{8\llbracket \epsilon_2\rrbracket\llbracket -\epsilon_1+\epsilon_2\rrbracket \prod_{\alpha=1}^N\llbracket \epsilon_1+\zeta\pm\mathfrak{a}_\alpha\rrbracket \llbracket \epsilon_1+\epsilon_2-\zeta\pm\mathfrak{a}_\alpha\rrbracket \llbracket 2\epsilon_1+\epsilon_2-\zeta\pm\mathfrak{a}_\alpha\rrbracket }\cr
\times\llbracket 2\zeta-2 \epsilon_1-\epsilon_2\rrbracket \llbracket 2\zeta-3\epsilon_1-\epsilon_2\rrbracket \llbracket -2\zeta+3\epsilon_1+2\epsilon_2\rrbracket~.
\end{align}

The origin of the factor $\llbracket -2\epsilon_+\rrbracket =\llbracket -(\epsilon_1+\epsilon_2)\rrbracket$ can be explained as follows.
As in \eqref{SO-def}, the defect contribution include the terms
\begin{equation}
    \cS\lt(\phi_i\pm \zeta\rt)= \frac{\llbracket \epsilon_- \pm(\phi_i\pm \zeta)\rrbracket }{\llbracket \epsilon_+\pm(\phi_i\pm \zeta)\rrbracket }~.
\end{equation}
However, when the pole $\phi_1= \zeta-\epsilon_+-\epsilon_1$ is taken, the following factor becomes
\begin{equation}
    \llbracket \epsilon_-+\phi_1- \zeta\rrbracket =\llbracket -(\epsilon_2+\epsilon_1)\rrbracket =\llbracket -2\epsilon_+\rrbracket ,
\end{equation}
in the numerator of $\cS(\phi_i\pm  \zeta)$. The astute reader may notice that there are also $\llbracket \pm 2\epsilon_+\rrbracket$ even in the denominator after evaluating the residue, but they cancel the prefactor $\llbracket  2\epsilon_+\rrbracket^k$ in \eqref{int-ZOe}. Consequently, the qq-character of SO($2N$) gauge group is not simply written in terms of $Y$ and $1/Y$ at the refined level \cite{Haouzi:2020yxy}. However, this pole yields a trivial residue in the unrefined limit $\e_+=0$.  A similar argument holds for the poles $\phi_i=\pm\zeta -\e_+-\e_{1,2}$ and $\phi_i=\pm\zeta +\e_++\e_{1,2}$. Their residues become trivial as they are proportional to $\llbracket \pm2\epsilon_+\rrbracket$ at the refined level. (Note that the residues at $\phi_i=\pm\zeta -\e_++\e_{1,2}$ and 
 $\phi_i=\pm\zeta +\e_+-\e_{1,2}$ are zero even at refine level.)

Note that this phenomenon can also be observed in the instanton partition function without a defect \cite{Nawata:2021dlk}. Poles that cannot be classified by Young diagrams provide residues proportional to $\llbracket \pm2\epsilon_+\rrbracket$. However, at the unrefined level, these non-trivial poles do not contribute to the partition function as their residues vanish.

The above argument applies to all $BCD$-type gauge theories as they all possess a similar structure, and for an arbitrary number of instantons.
In this manner, it can be shown that poles containing each defect parameter can be selected only once in each set of JK residues in the unrefined limit. Thus, this ensures the finiteness of qq-characters, which are expressed as finite combinations of expectation values of $Y$-operators in the unrefined limit.

\section{Vertical representations in the unrefined limit}\label{a:rep}

In this Appendix, we present the verification of the algebraic relations satisfied by the vertical  representations in the unrefined limit.

\subsection*{Type $A$}
The vertical representation in the unrefined limit is specified by \eqref{vrep-unref-A-ket}, and here we verify the commutation relations of the Drinfeld currents. In the vertical representations, we set $\hat{\ell}_1\mapsto 0$ (i.e. $\hat{\gamma}\mapsto 1$), hence all $h_m$'s commute. Therefore, the only non-trivial relations that need to be verified are those in \eqref{alg-unref-2}.

Let us show the first commutation relation of \eqref{alg-unref-2}. 
A useful equation is 
\begin{align}\label{p-p}
    p_m\lt(q^{(\lambda+x)^t+\rho+1/2}\rt)-p_m\lt(q^{\lambda^t+\rho+1/2}\rt)=&\lt\{q^{\lambda^t_j+1-(j-1)}\rt\}^m-\lt\{q^{\lambda^t_j-(j-1)}\rt\}^m\cr
    =&(q^{m}-1)\lt\{q^{\lambda^t_j+1-j}\rt\}^m,
\end{align}
Also, the delta function admits the following expansion
\be\label{deltafunctiondef}
    \delta(z)=\sum_{m\in\mathbb{Z}}z^{m}~.
\ee
Then, the explicit computation of the commutation relation is given by
\begin{align}
  &  \lt[h_m,x^+(z)\rt]\ket{\vec{A},\vec{\lambda}}\cr
    =&\frac{\llbracket \hbar\rrbracket}{m(1-q^{m})}\sum_{\alpha=1}^N \sum_{x\in \frakA(\vec{\lambda})}A^m_\alpha\delta(z/\chi_x)\lt(p_m\lt(q^{(\lambda^{(\alpha)}+x)^{t}+\rho+1/2}\rt)-p_m\lt(q^{(\lambda^{(\alpha)})^{t}+\rho+1/2}\rt)\rt)\ket{\vec{A},\vec{\lambda}+x}\cr
    =&-\frac{\llbracket \hbar\rrbracket}{m}\sum_{\alpha=1}^N \sum_{x\in\frakA(\lambda^{(\alpha)})}A_\alpha^m\lt\{q^{(\lambda^{(\alpha)})^{t}_j+1-j}\rt\}^m\delta(z/\chi_x)\ket{\vec{A},\vec{\lambda}+x}\cr
    =&-\frac{\llbracket \hbar\rrbracket }{m}\sum_{\alpha=1}^N \sum_{x\in\frakA(\lambda^{(\alpha)})}\chi_x^m\delta(z/\chi_x)\ket{\vec{A},\vec{\lambda}+x}\cr
    =&-\frac{\llbracket \hbar\rrbracket z^m}{m}\sum_{\alpha=1}^N \sum_{x\in\frakA(\lambda^{(\alpha)})}\delta(z/\chi_x)\ket{\vec{A},\vec{\lambda}+x}\cr
    =&-\frac{z^m}{m}x^+(z)\ket{\vec{A},\vec{\lambda}},
\end{align}
where $x=((\lambda^{(\alpha)})^{t}_j+1,j)\in \frakA(\lambda^{(\alpha)})$. One can show the commutation relation $\lt[h_m,x^-(z)\rt]$ in a similar way.

To show the second commutation relation of \eqref{alg-unref-2}, we notice the identity
\begin{align}\label{C4}
     &\sum_{m=1}^{\infty}m\llbracket\pm m\hbar\rrbracket h_mz^{-m}\ket{A,\lambda}=\sum_{m=1}^\infty m(1-q^m)(1-q^{-m})h_mz^{-m}\ket{A,\lambda}\nn\\
    =&\sum_{m=1}^\infty (1-q^{-m})\sum_{j=1}^\infty A^m \lt[q^{\lambda^t_j+1-j}\rt]^mz^{-m}\ket{A,\lambda}\\
        =&\lt[
\sum_{m=1}^\infty \sum_{j=1}^\infty A^m \lt[q^{\lambda^t_j+1-j}\rt]^mz^{-m}
-\sum_{m=1}^\infty \sum_{j=1}^\infty A^m \lt[q^{\lambda^t_j-j}\rt]^mz^{-m}
 \rt]\ket{A,\lambda},\nn\\
    =&\lt[\sum_{x\in \frakA(\lambda)}\sum_{m=1}^\infty (\chi_x/z)^m-\sum_{x\in \frakR(\lambda)}\sum_{m=1}^\infty (\chi_x/z)^m\rt]\ket{A,\lambda},\nonumber
\end{align}
which follows from the cancellation between columns satisfying $\lambda^t_j=\lambda^t_{j+1}$. Similarly, we also have the identity of the negative modes
\begin{align}\label{C5}
   \sum_{m=-1}^{-\infty}m\llbracket\pm m\hbar\rrbracket h_mz^{-m}\ket{A,\lambda}=&\sum_{m=-1}^{-\infty} m(1-q^m)(1-q^{-m})h_mz^{-m}\ket{A,\lambda}\\
    =&\sum_{m=1}^{\infty}(1-q^{m})\sum_{j=1}^\infty A^{-m} \lt[q^{\lambda^t_j+1-j}\rt]^{-m}z^{m}\ket{A,\lambda}\cr
    =&\lt[\sum_{x\in \frakA(\lambda)}\sum_{m=1}^\infty (\chi_x/z)^{-m}-\sum_{x\in \frakR(\lambda)}\sum_{m=1}^\infty (\chi_x/z)^{-m}\rt]\ket{A,\lambda}~.\nonumber
\end{align}
Then the explicit computation of the commutation relations is given by
\begin{align}
\lt[x^+(z),x^-(w)\rt]\ket{\vec{A},\vec{\lambda}}
  =&\llbracket \hbar\rrbracket^2\delta(z/w)\lt[\sum_{x\in \frakA(\vec{\lambda})}\delta(z/\chi_x)-\sum_{x\in \frakR(\vec{\lambda})}\delta(z/\chi_x)\rt]\ket{\vec{A},\vec{\lambda}}\cr
  =&-\llbracket \pm\hbar\rrbracket\delta(z/w)\lt[\sum_{m\neq 0}m\llbracket\pm m\hbar\rrbracket h_mz^{-m}+N\rt]\ket{\vec{A},\vec{\lambda}}\\
  =&-\llbracket \pm\hbar\rrbracket\lt[\delta(z/w)\sum_{m\neq 0}m\llbracket\pm m\hbar\rrbracket h_mz^{-m}+N\sum_{n\in\mathbb{Z}}(z/w)^n\rt] \ket{\vec{A},\vec{\lambda}},\nonumber
\end{align}
which shows the second relation of \eqref{alg-unref-2} with $\hat{\ell}_1\mapsto 0$ and $\hat{\ell}_2\mapsto N$. The first line results from the fact that the nontrivial parts only come from adding or removing the same boxes from the Young diagrams.

In conclusion, \eqref{vrep-unref-A-ket} is a representation of the unrefined limit \eqref{alg-unref-2} of the quantum toroidal algebra $\QTA$.

\subsection*{Type $BD$}

Now let us move on to the $BD$-type algebra \eqref{BCD-alg}, and verify the validity of the vector representations \eqref{BCD-vert-rep}. The method is analogous to the $A$-type.
Using \eqref{p-p}, we can show the first commutation relation of  \eqref{BCD-alg}
\begin{align}
    &\lt[h_m,x^+(z)\rt]\ket{\vec{A},\vec{\lambda}}\cr
    =&-{\llbracket \hbar\rrbracket}\sum_{x\in \frakA(\vec{\lambda})}\lt(\delta(z/\chi_x)\lt(\chi_x^{m}+\chi_x^{-m}\rt)+\delta(z\chi_x)\lt(\chi_x^{m}+\chi_x^{-m}\rt)\rt)\ket{\vec{A},\vec{\lambda}+x}\cr
    =&-(z^{m}+z^{-m})x^+(z)\ket{\vec{A},\vec{\lambda}},
\end{align}
 One can verify the commutation relation $\lt[h_m,x^-(z)\rt]$ in a similar way.
 
Similar to \eqref{C4} and \eqref{C5}, we have
\begin{align}
\sum_{\substack{m\in\mathbb{Z}\\m\neq 0}}\llbracket\pm m\hbar\rrbracket h_mz^{-m}\ket{A,\lambda}=&\sum_{\substack{m\in\mathbb{Z}\\m\neq 0}} (1-q^{m})(1-q^{-m})h_mz^{-m}\ket{A,\lambda}\cr
    =&\sum_{\substack{m\in\mathbb{Z}\\m\neq 0}} z^{-m} (1-q^{-m})\sum_{j=1}^\infty A^{m} \lt(q^{\lambda^t_j-(j-1)}\rt)^{m}\\
    &\qquad +z^{-m} (1-q^{m})\sum_{j=1}^\infty A^{-m} \lt(q^{\lambda^t_j-(j-1)}\rt)^{-m}\ket{A,\lambda}\cr
    =&\sum_{\substack{m\in\mathbb{Z}\\m\neq 0}}\lt[\bigg(\sum_{x\in \frakA(\lambda)}-\sum_{x\in \frakR(\lambda)}\bigg)(z^{-m}\chi_x^{m}+z^{-m}\chi_x^{-m})\rt]\ket{A,\lambda}\cr
    =&\lt[\bigg(\sum_{x\in \frakA(\lambda)}-\sum_{x\in \frakR(\lambda)}\bigg)(\delta(z/\chi_x)+\delta(z\chi_x)-2)\rt]\ket{A,\lambda}\nonumber
\end{align}
Then, the second commutation relation of  \eqref{BCD-alg}  with $\hat{\ell}_1\mapsto 0$ and $\hat{\ell}_2\mapsto N$ follows:
\begin{align}
  &  \lt[x^+(z),x^-(w)\rt]\ket{\vec{A},\vec{\lambda}}\cr
=&\llbracket \hbar\rrbracket^2\bigg[\sum_{x\in \frakA(\vec{\lambda})}\big(\delta(z/w)\delta(z/\chi_x)+\delta(zw)\delta(z/\chi_x)+\delta(zw)\delta(z\chi_x)+\delta(z/w)\delta(z\chi_x)\big)\cr
&-\sum_{x\in \frakR(\vec{\lambda})}\big(\delta(z/w)\delta(z/\chi_x)+\delta(zw)\delta(z/\chi_x)+\delta(zw)\delta(z\chi_x)+\delta(z/w)\delta(z\chi_x)\big)\bigg]\ket{\vec{A},\vec{\lambda}}\cr
    =&\llbracket \hbar\rrbracket^2\lt(\delta(z/w)+\delta(zw)\rt)\Big[\big(\sum_{x\in \frakA(\vec{\lambda})}-\sum_{x\in \frakR(\vec{\lambda})}\big)\lt(\delta(z/\chi_x)+\delta(z\chi_x)\rt)\Big]\ket{\vec{A},\vec{\lambda}}\cr
    =&-\llbracket \pm\hbar\rrbracket\lt(\delta(z/w)+\delta(zw)\rt)\Big[\sum_{m\neq 0}\llbracket\pm m\hbar\rrbracket h_mz^{-m}+2N\Big]\ket{\vec{A},\vec{\lambda}}.
\end{align}
In conclusion, \eqref{BCD-vert-rep} is a representation of the $BD$-type algebra \eqref{BCD-alg}.

\section{Recursion relations of instanton partition functions}\label{a:id}

In this Appendix, we derive the recursion relations for the $\SU(N)$ and $\SO(n)$ instanton partition functions in terms of Young diagrams. These recursion relations are essential for the Ward identity approach to qq-characters in \S\ref{s:Ward}. To this end, we provide explicit expressions for the unrefined instanton partition functions as summations over Young diagrams and discuss their properties briefly.

\subsection*{Type $A$}

Let us first consider the recursion relation of the instanton partition function \eqref{instnaton-A} for $A$-type gauge group.
 In this paper, several different types of $Y$-functions are introduced to give analytic expressions to qq-characters. These $Y$-functions are all related to the $A$-type $Y$-function \eqref{def-YA}. The eigenvalue of the $Y^A$-operator (\ref{YA}) on the state $\ket{\vec{A},\vec{\lambda}}$ can be factorized into a product of $\cY^A$-functions, each depending solely on a single Young diagram:
\begin{equation}
    \cY^A_{\vec{\lambda}}(z)=\prod_{\alpha=1}^N\cY^A_{\lambda^{(\alpha)}}(z),
\end{equation}
where the $\cY$-function associated with a single Young tableau $\lambda$ and Coulomb branch parameter $\mathfrak{a}$ is given by
\begin{equation}
\label{Yex}
\cY^A_{\lambda}(z)=\llbracket \zeta-\mathfrak{a}+2\e_+\rrbracket \prod_{x\in\lambda}\cS(\zeta-\phi_x)~.
\end{equation}
For this reason, it suffices to enumerate the various useful properties of the $\cY$-function for a single Young tableau. The most important formula, referred to as the ``shell formula'' in some literature, is:
\begin{align}
\cY^A_{\lambda}(z)=\frac{\prod_{x\in \frakA({\lambda})}\llbracket \zeta-\phi_x+\epsilon_+\rrbracket}{\prod_{x\in \frakR({\lambda})}\llbracket \zeta-\phi_x-\epsilon_+\rrbracket}.
\end{align}
Another useful expression that is easy to derive from the definition of $\cY$-function is
\begin{align}
    \cY^A_\lambda(z)=&-\lt(\frac{zq_1q_2}{A}\rt)^{\frac{1}{2}}\prod_{x'\in{\cal X}_\lambda}\frac{1-\frac{x'}{zq_1q_2}}{1-\frac{q_2x'}{zq_1q_2}}\label{y-x-1}\\
    =&\exp\lt(-\sum_{m,j=1}^\infty \frac{1}{m}(1-q_2^m)A^mq_2^{m(j-2)}q_1^{m(\lambda^t_j-1)}z^{-m}+\pi i+\frac{1}{2}\log(zq_1q_2/A)\rt),\nonumber
\end{align}
and equivalently
\begin{align}
    \cY^A_\lambda(z)=&\lt(\frac{A}{zq_1q_2}\rt)^{\frac{1}{2}}\prod_{x'\in{\cal X}_\lambda}\frac{1-\frac{zq_1q_2}{x'}}{1-\frac{zq_1q_2}{q_2x'}}\label{y-x-2}\\
    =&\exp\lt(-\sum_{m,j=1}^\infty \frac{1}{m}\sum_k(1-q_2^{-m})A^{-m}q_2^{-m(j-2)}q_1^{-m(\lambda^t_j-1)}z^{m}+\frac{1}{2}\log(A/zq_1q_2)\rt),\nonumber
\end{align}
where ${\cal X}_\lambda=\{Aq_2^{j-1}q_1^{\lambda^t_j}\}_{j=1}^\infty$.

From now on, we consider only the unrefined limit, and the unrefined limit of the Nekrasov factor \eqref{def:Nekra} admits an equivalent expression and associated recursive formulae \cite{Awata:2008ed}:
\begin{equation}
    N_{\lambda\nu}(Q,q)=\prod_{i,j=1}^\infty\frac{1-Qq^{-\nu^t_j-\lambda_i+i+j-1}}{1-Qq^{i+j-1}}.\label{Nekra-IKV}
\end{equation}
It then follows that
\begin{align}
    \frac{N_{(\lambda^{(1)}+x)\lambda^{(2)}}(A_1/A_2;q)}{N_{\lambda^{(1)}\lambda^{(2)}}(A_1/A_2;q)}=&\prod_{x^\prime\in \cX_{\lambda^{(2)}}}\frac{1-\chi_x/x^\prime}{1-q\chi_x/x^\prime}=(\chi_x/A_2)^{\frac12}\cY^A_{\lambda^{(2)}}(\chi_x),\cr
\frac{N_{\lambda^{(1)}(\lambda^{(2)}+x)}(A_1/A_2;q)}{N_{\lambda^{(1)}\lambda^{(2)}}(A_1/A_2;q)}=&\prod_{x^\prime\in \cX_{\lambda^{(1)}}}\frac{1-x^\prime/\chi_x}{1-x^\prime/q\chi_x}=-(A_1/\chi_x)^{\frac12}\cY^A_{\lambda^{(1)}}(\chi_x),\label{Nekra-recursive-Y2}
\end{align}
which can alternatively be checked directly with the \textit{Mathematica} file attached to this paper.
Combining the above two recursive relations together, we obtain
\begin{align}
    \frac{N_{(\lambda+x)(\lambda+x)}(1;q)}{N_{\lambda\lambda}(1;q)}=&-\lim_{z\rightarrow \chi_x}\cY^A_{\lambda+x}(z)\cY^A_\lambda(z)\cr
    =&-\llbracket \pm\hbar\rrbracket\lim_{z\rightarrow \chi_x}\llbracket \zeta-\phi_x\rrbracket ^{-2}\lt(\cY^A_\lambda(z)\rt)^2\cr
    =&-\llbracket \pm\hbar\rrbracket^{-1}\lim_{z\rightarrow \chi_x}\llbracket \zeta-\phi_x\rrbracket ^{2}\lt(\cY^A_{\lambda+x}(z)\rt)^2.\label{Nekra-rec-2}
\end{align}
Another important identity we need in the rewriting of the instanton partition functions is
\begin{equation}
N_{\lambda\nu}(Q;q)=(-Q)^{|\lambda|+|\nu|}q^{\frac{1}{2}\kappa(\lambda)-\frac{1}{2}\kappa(\nu)}N_{\nu\lambda}\lt(Q^{-1};q\rt),\label{Nekra-convert-formu}
\end{equation}

Let us recall that the pure SU($N$) Nekrasov instanton partition function \eqref{instnaton-A}  can be written as
\begin{equation}
Z_{\textrm{inst}}^{\fraksu(N)}=\sum_{\vec{\lambda}}\mathfrak{q}^{|\vec{\lambda}|}Z^{\fraksu(N)}_{\vec{\lambda}},
\end{equation}
where
\begin{equation}
Z^{\fraksu(N)}_{\vec{\lambda}}=\prod_{\alpha=1}^NN^{-1}_{\lambda^{(\alpha)}\lambda^{(\alpha)}}(1,q)\prod_{\alpha\neq \beta}N^{-1}_{\lambda^{(\alpha)}\lambda^{(\beta)}}(A_\alpha/A_\beta,q)~.
\end{equation}
Using \eqref{Nekra-rec-2}, we can derive the recursive relations for $\lambda^{(\alpha)}\to \lambda^{(\alpha)}+x$ as follows:
\begin{align}
\frac{Z^{\fraksu(N)}_{\vec{\lambda}+x}}{Z^{\fraksu(N)}_{\vec{\lambda}}}=&-\frac{1}{\llbracket \pm\hbar\rrbracket}\lim_{\zeta\rightarrow\phi_x}\frac{\llbracket \zeta-\phi_x\rrbracket ^2}{\lt(\cY^A_{\lambda^{(\alpha)}}(z)\rt)^2}\prod_{\beta\neq\alpha}\lt(-\frac{1}{\lt(\cY^A_{\lambda^{(\beta)}}(\chi_x)\rt)^2}\rt)\cr
=&\frac{(-1)^{N}}{\llbracket \pm\hbar\rrbracket}\lim_{\zeta\rightarrow\phi_x}\frac{\llbracket \zeta-\phi_x\rrbracket ^2}{\lt(\cY^A_{\vec{\lambda}}(z)\rt)^2}\cr
=&(-1)^{N}\llbracket\pm \hbar\rrbracket\lim_{\zeta\rightarrow\phi_x}\frac{1}{\llbracket \zeta-\phi_x\rrbracket ^2\lt(\cY^A_{\vec{\lambda}+x}(z)\rt)^2}.\label{ZU-rec-2}
\end{align}

\subsection*{Type $BD$}
Now let us move on to $\SO(n)$ gauge groups.
The unrefined instanton partition function for the $\SO(2N)$ gauge group can be written as a summation over Young diagrams \cite{Nawata:2021dlk}. The partition function is given by:
\begin{align}\label{Z-SOeven-r}
Z^{\frakso(2N)}_{\textrm{inst}}=&\sum_{\vec{\lambda}}\mathfrak{q}^{|\vec{\lambda}|} Z^{\frakso(2N)}_{\vec{\lambda}}\cr :=&\sum_{\vec{\lambda}}\mathfrak{q}^{|\vec{\lambda}|} M_{\vec{\lambda}}(\vec{A})^2\prod_{\alpha=1}^NA_\alpha^{(N-4)|\lambda^{(\alpha)}|}\prod_{\alpha=1}^N q^{-\frac{(N-4)\kappa(\lambda^{(\alpha)})}{2}}
   \cr
& \times \prod_{\alpha=1}^NA_\alpha^{|\vec{\lambda}|}N_{\lambda^{(\alpha)}\lambda^{(\alpha)}}(1,q)^{-1}\prod_{\alpha\neq \beta}  N_{\lambda^{(\alpha)}\lambda^{(\beta)}}(A_{\alpha}A_\beta^{-1},q)^{-1}
\end{align}
where $\kappa(\lambda):=2 \sum_{(i, j) \in \lambda}(j-i)$, and the $M$-factor is expressed as
\begin{align}\label{product}
&M_{\vec{\lambda}}(A_1,A_2,\dots,A_N)\cr:=&\frac{\displaystyle\prod_{\alpha=1}^N\prod_{(i,j)\in\lambda^{(\alpha)}}(1-A_{\alpha}^{2}q^{i-j+(\lambda^{(\alpha)})_{j}^t-\lambda_{i}^{(\alpha)}})}
{\displaystyle\prod_{1\le \alpha<\beta\le N}\prod_{(i,j)\in\lambda^{(\alpha)}}(1-A_{\alpha} A_{\beta} q^{i+j-1-\lambda^{(\alpha)}_{i}-\lambda^{(\beta)}_{j}})\prod_{(m,n)\in\lambda^{(\beta)}}(1-A_{\alpha} A_{\beta} q^{1-m-n+(\lambda^{(\beta)})^t_{n}+(\lambda^{(\alpha)})^t_{m}})}~.\nonumber
\end{align}
Hence, to derive the recursion relations for the partition function, it is necessary to derive the recursive formula for the $M$-factor by adding a box in the Young diagrams $\lambda^{(\alpha)}\to \lambda^{(\alpha)}+x$
\begin{equation}
\frac{M_{\vec{\lambda}+x}}{M_{\vec{\lambda}}}=\textrm{P.E.}(X),\label{rec-def-M}
\end{equation}
with $x\in (m,\lambda_m^{(\a)}+1)$  and
\begin{align}
    X(\vec{\lambda})=&-A_{\alpha}^{2}q^{m-1-\lambda_m^{(\alpha)}}\Bigl( (1+q)q^{m-1-\lambda_m^{(\alpha)}}-q^{\ell(\lambda^{(\alpha)})} +(q-1)\sum_{i\neq m}q^{i-1-\lambda_i^{(\alpha)}} \Bigr)  \cr
&+A_{\alpha}q^{m-1-\lambda_m^{(\alpha)}}\sum_{\beta\neq \alpha}A_{\beta} \Bigl(  q^{\ell(\lambda^{(\beta)})}+(1-q)\sum_{i=1}^{\ell(\lambda^{(\beta)})}q^{i-1-\lambda^{(\beta)}_i}\Bigr) .
\end{align}
Using $\chi_x=A_\alpha q^{m-\lambda^{(\alpha)}_m-1}$, we can rewrite
\begin{align}
    X(\vec{\lambda})=&-\chi_x\lt((1+q)\chi_x-A_\alpha q^{\ell(\lambda^{(\alpha)})}+(1-q^{-1})\sum_{y\in \frakS(\lambda^{(\alpha)}),y\neq x-(0,1)}\chi_y\rt)\cr
   & +\chi_x\sum_{\beta\neq \alpha}\lt(A_\beta q^{\ell(\lambda^{(\beta)})}-(1-q^{-1})\sum_{y\in \frakS(\lambda^{(\beta)})}\chi_y\rt),
\end{align}
where we denoted $\frakS(\lambda)=\{(i,\lambda_i)\}_{i=1}^{\ell(\lambda)}$ as the set of boxes at the surface of a Young diagram $\lambda$. Note that if $\lambda_i=\lambda_{i+1}$ then $\chi_{(i,\lambda_i)}=q^{-1}\chi_{(i+1,\lambda_{i+1})}$, and thus
\begin{align}
    X(\vec{\lambda})=&-\chi_x\lt(\chi_x+\sum_{y\in \frakR(\lambda^{(\alpha)})}\chi_y-\sum_{y\in \frakA(\lambda^{(\alpha)}),y\neq x}\chi_y\rt)\cr
   & -\chi_x\sum_{\beta\neq \alpha}\lt(\sum_{y\in \frakR(\lambda^{(\beta)})}\chi_y-\sum_{y\in \frakA(\lambda^{(\beta)})}\chi_y\rt).
\end{align}
Substituting this back to \eqref{rec-def-M}, we have the following recursive relation
\begin{equation}
    \frac{M_{\vec{\lambda}+x}}{M_{\vec{\lambda}}}=(1-\chi_x^2)\prod_{\beta=1}^N\frac{\prod_{y\in \frakR(\lambda^{(\beta)})}(1-\chi_x\chi_y)}{\prod_{\substack{y\in \frakA(\lambda^{(\beta)})\\y\neq x}}(1-\chi_x\chi_y)}.
\end{equation}
One can directly test the above formula with the \textit{Mathematica} file attached to this paper on the \textit{arXiv}.

Note that
\begin{equation}
q^{c\kappa(\lambda+x)/2}/q^{c\kappa(\lambda)/2}=\chi_x^{-c}A^{c},
\end{equation}
we obtain the following recursive relation for the SO($2N$) partition function \eqref{Z-SOeven-r}
\begin{align}
\frac{Z^{\frakso(2N)}_{\vec{\lambda}+x}}{Z^{\frakso(2N)}_{\vec{\lambda}}}=\lt(\prod_{\alpha=1}^NA_\alpha\rt)\chi_x^{(N-4)}\frac{(-1)^{N-1}}{\llbracket \hbar\rrbracket^2}\lim_{\zeta\rightarrow\phi_x}\frac{\llbracket \zeta-\phi_x\rrbracket ^2}{\lt(\cY^A_{\vec{\lambda}}(z)\rt)^2}\cr
    \times\lt((1-\chi_x^2)^2\prod_{\beta=1}^N\frac{\prod_{y\in \frakR(\lambda^{(\beta)})}(1-\chi_x\chi_y)}{\prod_{y\in \frakA(\lambda^{(\beta)})}(1-\chi_x\chi_y)}\rt)^2.
\end{align}
Using the $D$-type $\cY$-function defined in \eqref{def-YBD} and its shell formula
\begin{equation}
\cY^D_{\vec\lambda}(z)=\frac{\prod_{x\in \frakA(\vec{\lambda})}\llbracket \zeta\pm\phi_x\rrbracket }{\prod_{x\in \frakR(\vec{\lambda})}\llbracket \zeta\pm\phi_x\rrbracket }=-(z^N\prod_{\alpha=1}^NA_\alpha)^{-\frac12}\cY^A_{\vec\lambda}(z)\frac{\prod_{x\in \frakA(\vec{\lambda})}(1-\chi_xz)}{\prod_{x\in \frakR(\vec{\lambda})}(1-\chi_x z)},\label{YD-pole}
\end{equation}
we arrive at the recursive formulae written in terms of the $\cY^D$-function,
\begin{align}
    \frac{Z^{\frakso(2N)}_{\vec{\lambda}+x}}{Z^{\frakso(2N)}_{\vec{\lambda}}}=&\frac{1}{\llbracket \hbar\rrbracket^2}\lim_{\zeta\rightarrow\phi_x}\frac{\llbracket \zeta-\phi_x\rrbracket ^2\llbracket 2\zeta\rrbracket ^4}{\lt(\cY^D_{\vec{\lambda}}(z)\rt)^2}\cr
     =&(-\llbracket \hbar\rrbracket)^2\lim_{\zeta\rightarrow\phi_x}\frac{\llbracket 2\zeta\pm\hbar\rrbracket ^2}{\llbracket \zeta-\phi_x\rrbracket ^2\lt(\cY^D_{\vec{\lambda}+x}(z)\rt)^2}.\label{Z-rec-D}
\end{align}

For SO($2N+1$) theory, the partition function is given by 
\begin{align}\label{Z-SOodd-r}
Z^{\frakso(2N+1)}_{\textrm{inst}}=&\sum_{\vec{\lambda}}\mathfrak{q}^{|\vec{\lambda}|} Z^{\frakso(2N+1)}_{\vec{\lambda}}\\ :=&\sum_{\vec{\lambda}}\mathfrak{q}^{|\vec{\lambda}|} M_{\vec{\lambda}}(\vec{A})^2\prod_{\alpha=1}^NA_\alpha^{(N-4)|\lambda^{(\alpha)}|}\prod_{\alpha=1}^N (-q)^{-\frac{(N-4)\kappa(\lambda^{(\alpha)})}{2}}
  \prod_{\alpha\neq \beta}  N_{\lambda^{(\alpha)}\lambda^{(\beta)}}(A_{\alpha}A_\beta^{-1},q)^{-1} \cr
& \times \prod_{\alpha=1}^NA_\alpha^{|\vec{\lambda}|}N_{\lambda^{(\alpha)}\lambda^{(\alpha)}}(1,q)^{-1}N_{\lambda^{(\alpha)}\emptyset}(A_\a,q)^{-1}N_{\emptyset\lambda^{(\alpha)}}(A_\a^{-1},q)^{-1}~.\nonumber
\end{align}
The key distinction between \eqref{Z-SOeven-r} and \eqref{Z-SOodd-r} is the presence of the terms $N_{\lambda^{(\alpha)}\emptyset}N_{\emptyset\lambda^{(\alpha)}}$, whose recursion relation is 
\begin{equation}
\frac{N_{\lambda^{(\alpha)}+x\emptyset}(A_\a,q)N_{\emptyset\lambda^{(\alpha)}+x}(A_\a^{-1},q)}{N_{\lambda^{(\alpha)}\emptyset}(A_\a,q)N_{\emptyset\lambda^{(\alpha)}}(A_\a^{-1},q)}=-\llbracket \chi_x\rrbracket^2
\end{equation}
This recursion relation compensates for the difference in $\cY$-functions \eqref{def-YBD} between the $\SO(2N)$ and $\SO(2N+1)$ gauge groups, as given by
$$\big(\cY^{B_N}_\lambda(z)\big)^2=\llbracket \zeta\rrbracket^2 \big( \cY^{D_N}_\lambda(z)\big)^2~.$$
In summary, the recursive relation for the $B$-type instanton partition functions is
\begin{align}
     \frac{Z^{\frakso(2N+1)}_{\vec{\lambda}+x}}{Z^{\frakso(2N+1)}_{\vec{\lambda}}}=&\frac{1}{\llbracket \hbar\rrbracket^2}\lim_{\zeta\rightarrow\phi_x}\frac{\llbracket \zeta-\phi_x\rrbracket ^2\llbracket 2\zeta\rrbracket ^4}{\llbracket \zeta\rrbracket ^2\lt(\cY^D_{\vec{\lambda}}(z)\rt)^2}=\frac{1}{\llbracket \hbar\rrbracket^2}\lim_{\zeta\rightarrow\phi_x}\frac{\llbracket \zeta-\phi_x\rrbracket ^2\llbracket 2\zeta\rrbracket ^4}{\lt(\cY^B_{\vec{\lambda}}(z)\rt)^2}\cr
     =&\llbracket \hbar\rrbracket^2\lim_{\zeta\rightarrow\phi_x}\frac{\llbracket 2\zeta\pm\hbar\rrbracket ^2}{\llbracket \zeta-\phi_x\rrbracket ^2\lt(\cY^B_{\vec{\lambda}+x}(z)\rt)^2}.\label{Z-rec-B}
\end{align}
The formulas \eqref{Z-rec-D} and \eqref{Z-rec-B} can again be checked with our \textit{Mathematica} file.

\subsection{proof of pole cancellation}\label{app:pole-cancel}

Now we are ready to prove the pole cancellations in the expressions of qq-characters \eqref{BD-qq}. For the $C$-type, we offer a partial answer to the pole cancellation of \eqref{C-qq-pre}.

Recall from the expression \eqref{YD-pole} and the definitions \eqref{def-YBD} and \eqref{Y-C} that 
\begin{align}
    &\cY^{B_N}_{\vec{\lambda}}(z)=\llbracket \zeta\rrbracket\cY^{D_N}_{\vec{\lambda}}(z),\\
    &\cY^{C_N}_{\vec{\lambda}}(z)=\frac{1}{\llbracket 2\zeta\rrbracket^2\llbracket 2\zeta\pm \hbar\rrbracket}\cY^{D_{N+4}}_{\vec{\lambda}}(z),\label{YCinD}
\end{align}
so there are poles at $\zeta=\phi_{x\in \mathfrak{R}(\vec{\lambda})}$ in $\cY^{BCD}_{\vec{\lambda}}(z)$ and poles at $\zeta=\phi_{x\in \mathfrak{A}(\vec{\lambda})}(z)$ in $1/\cY^{BCD}_{\vec{\lambda}}(z)$. The set of such poles from $\cY^{BCD}_{\vec{\lambda}}(z)$ with all Young diagrams of size $k+1$ coincide with that of $1/\cY^{BCD}_{\vec{\lambda}}(z)$ with Young diagram size $k$. We also notice that although there are four sectors in the Sp($N$) calculation, the relation \eqref{YCinD} holds universally for all the sectors. 

Let us first evaluate the residue of $\cY^{D_N}_{\vec{\lambda}+x}(z)Z^{\frakso(2N)}_{\vec{\lambda}+x}$ at $\zeta=\phi_x$, which is contained in the expectation value $\langle Y^D(z)\rangle $. 
\begin{align}
    \lim_{\zeta\rightarrow \phi_x}\llbracket \zeta-\phi_x\rrbracket\cY^{D_N}_{\vec{\lambda}+x}(z)Z^{\frakso(2N)}_{\vec{\lambda}+x}=&\lim_{\zeta\rightarrow \phi_x}\llbracket \zeta-\phi_x\rrbracket \frac{Z^{\frakso(2N)}_{\vec{\lambda}+x}}{Z^{\frakso(2N)}_{\vec{\lambda}}}\cY^{D_N}_{\vec{\lambda}+x}(z)Z^{\frakso(2N)}_{\vec{\lambda}}\cr
    =&\lim_{\zeta\rightarrow \phi_x}\llbracket \zeta-\phi_x\rrbracket \frac{\llbracket \hbar\rrbracket^2\llbracket 2\zeta\pm\hbar\rrbracket ^2}{\llbracket \zeta-\phi_x\rrbracket ^2\cY^D_{\vec{\lambda}+x}(z)}Z^{\frakso(2N)}_{\vec{\lambda}}\cr
    =&-\lim_{\zeta\rightarrow \phi_x}\llbracket \zeta-\phi_x\rrbracket \frac{\llbracket2\zeta\rrbracket^2\llbracket 2\zeta\pm\hbar\rrbracket}{\cY^D_{\vec{\lambda}}(z)}Z^{\frakso(2N)}_{\vec{\lambda}},
\end{align}
where we use the recursive relation \eqref{Z-rec-D} and \eqref{def-YBD}, \eqref{S} in the unrefined limit. The above rewriting shows that the residue at $\zeta=\phi_x$ from $\langle Y^D(z)\rangle$ is precisely equal to the negative of the residue at $\zeta=\phi_x$ from $\langle \frac{\llbracket2\zeta\rrbracket^2\llbracket 2\zeta\pm\hbar\rrbracket}{Y^D(z)}\rangle$. This shows that the pole is cancelled and the overall net residue at $\zeta=\phi_x$ in the qq-character \eqref{BD-qq} is zero. The same argument can be applied analogously in the $B$-type case. 

\begin{figure}[ht]
    \centering
    \includegraphics[width=5cm]{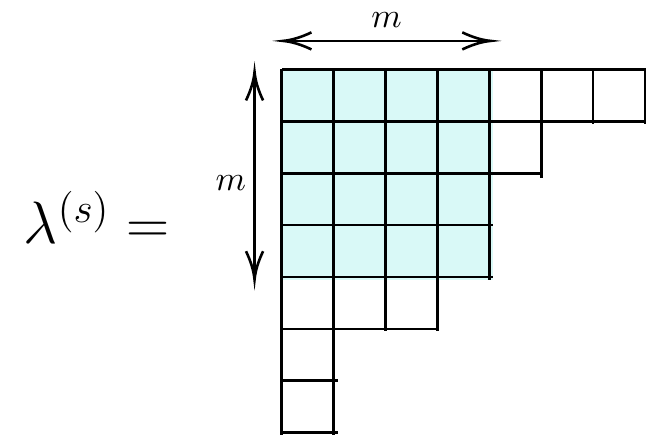}
    \caption{For $A_s=\pm 1,\pm q^{\frac{1}{2}}$, $m$ is defined to count the number of rows satisfying $\lambda^{(s)}_j\geq j$. }
    \label{fig:factor-1}
\end{figure}

\begin{figure}[ht]
    \centering
    \includegraphics[width=5cm]{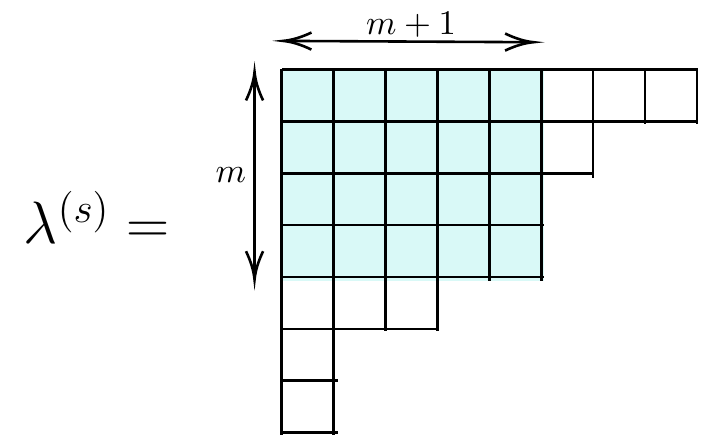}
    \caption{For $A_s=\pm q$, $m$ is defined to count the number of rows satisfying $\lambda^{(s)}_j\geq j+1$.}
    \label{fig:factor-2}
\end{figure}

The computation for the $C$-type case follows a similar pattern, but we need to incorporate the relation \eqref{Z-Sp}:
\begin{equation}
    \frac{Z^{\mathfrak{sp}(N),\pm}_{\vec{\lambda}+x}}{Z^{\mathfrak{sp}(N),\pm}_{\vec{\lambda}}}=\frac{C_{\vec{\lambda}+x,\vec{A}}}{C_{\vec{\lambda},\vec{A}}}\frac{Z^{\mathfrak{so}(2N+8)}_{\vec{\lambda}+x}}{Z^{\mathfrak{so}(2N+8)}_{\vec{\lambda}}},
\end{equation}
where four additional parameters $A_{s}$, $s=N+1,\dots,N+4$, are determined accordingly to the $\pm$-sectors and even/odd number of instantons and they take value in $\{\pm 1,\pm q^{\frac{1}{2}},\pm q\}$ \cite{Nawata:2021dlk}. The coefficients $C_{\vec{\lambda},\vec A}$ only depend on the number $m_s$ that counts the number of rows satisfying $\lambda^{(s)}_j\geq j$ or $\lambda^{(s)}_j\geq j+1$ in the extra Young diagrams labeled by $s=N+1,\dots,N+4$ depending on the value of $A_s$ (see Figure \ref{fig:factor-1} and \ref{fig:factor-2}). We call the boxes not at the specific positions (respectively with $x={}^\exists(i,i)$ and $x={}^\exists(i,i+1)$ in $\lambda^{(s)}$ for $s=N+1,\dots,N+4$) that change the characteristic number $m$, as generic boxes. Clearly, when the added box $x$ is a generic box, we have
\begin{equation}
    \frac{C_{\vec{\lambda}+x,\vec{A}}}{C_{\vec{\lambda},\vec{A}}}=1.
\end{equation}
For such generic boxes, the computation simplifies as follows
\begin{align}
    \lim_{\zeta\rightarrow \phi_x}\llbracket \zeta-\phi_x\rrbracket\cY^{C_N}_{\vec{\lambda}+x}(z)Z^{\fraksp(2N),\pm}_{\vec{\lambda}+x}=&\lim_{\zeta\rightarrow \phi_x}\llbracket \zeta-\phi_x\rrbracket \frac{Z^{\frakso(2N+8)}_{\vec{\lambda}+x}}{Z^{\frakso(2N+8)}_{\vec{\lambda}}}\frac{\cY^{D_{N+4}}_{\vec{\lambda}+x}(z)Z^{\fraksp(2N),\pm}_{\vec{\lambda}}}{\llbracket 2\zeta\rrbracket^2\llbracket 2\zeta\pm \hbar\rrbracket}\cr
    =&-\lim_{\zeta\rightarrow \phi_x}\llbracket \zeta-\phi_x\rrbracket \frac{1}{\cY^{D_{N+4}}_{\vec{\lambda}}(z)}Z^{\fraksp(2N),\pm}_{\vec{\lambda}}\\
    =&-\lim_{\zeta\rightarrow \phi_x}\llbracket \zeta-\phi_x\rrbracket \frac{1}{\llbracket2\zeta\rrbracket^2\llbracket 2\zeta\pm\hbar\rrbracket\cY^{C_{N}}_{\vec{\lambda}}(z)}Z^{\fraksp(2N),\pm}_{\vec{\lambda}}.\nonumber
\end{align}
The above calculation then shows the pole cancellation of the qq-character \eqref{C-qq} at a pole $\phi_x$ with a generic box $x$ of the extra Young diagrams. The poles corresponding to boxes on the special position are located at $\zeta=0,\pi i,\pm\frac{\hbar}{2},\pm(\frac{\hbar}{2}+\pi i),\pm\hbar,\pm(\hbar+\pi i)$. Interestingly we still observe highly non-trivial cancellations at these poles from instanton expansions up to six instantons. 

We remark that the pole cancellation shown here is an independent computation of that presented in Appendix \ref{a:qq}, but they both suggest the same form of the qq-characters schematically as $Y+1/Y$. 

\bibliographystyle{JHEP}
\bibliography{qq-toroidal}

\end{document}